\documentclass{elsart}
\usepackage{color,amsmath,amsfonts,amssymb}
 
\usepackage[dvips]{graphics}

\usepackage{mark_newcommands}

\usepackage{mark_widepage}
\usepackage{lscape}

\usepackage{mark_journals}

\begin{document} 
\begin{frontmatter}
\title{Efficient form of the
LANS-alpha turbulence model in a primitive-equation ocean model
}
\author{Mark R. Petersen}
\address{Computer, Computational and Statistical Sciences Division,
 and Center for Nonlinear Studies,
 Los Alamos National Laboratory, Los Alamos, New Mexico}
\ead{mpetersen@lanl.gov}

\begin{abstract}
The Lagrangian-Averaged Navier-Stokes alpha (LANS-$\alpha$) model is a
turbulence parameterization that has been shown to capture some of the
most important features of high resolution ocean modeling at lower
resolution.  Simulations using LANS-$\alpha$ in the POP
primitive-equation ocean model resemble doubled-resolution simulations
of standard POP in statistics like kinetic energy, eddy kinetic
energy, and potential temperature fields.  The computational cost of
adding LANS-$\alpha$ is only 27\% for our most efficient
implementation, as compared to a factor of 8---10 for a doubling of
resolution.

The LANS-$\alpha$ model improves turbulence statistics with an
additional nonlinear term and a smoothed advecting velocity.  In this
work we investigate different kinds of smoothing techniques and their
effect on the LANS-$\alpha$ model's results and efficiency. We show that
we can substitute convolution filters for full Hemlholtz inversions
and produce similar results at a significantly lower expense. 

When constructing filters for LANS-$\alpha$ in a primitive-equation
ocean model the filter weights must be chosen carefully, otherwise a
pressure-velocity instability will be excited.  We show analytically
that certain ranges of filter weights are unstable, and confirm this
with numerical experiments.  Our stability criterion also guarantees
that the kinetic energy is well defined, and that the filtered
velocity is smoother that the original velocity.  
\end{abstract}
\end{frontmatter}
\section{Introduction \label{s_introduction}}

The LANS-$\alpha$ model is a turbulence model that can reproduce some
of the statistics and structures that are seen in standard simulations
(without LANS-$\alpha$) at higher resolution.  The LANS-$\alpha$ model
accomplishes this in part by using a smoothed velocity field for the
advecting velocity in the momentum and tracer equations.  The question
addressed in this paper is, ``What is the best way to smooth the
velocity field in LANS-$\alpha$?''  This is not simply a matter of
choosing the smoothing method that produces the best result.  As a
turbulence model, LANS-$\alpha$ must be extremely efficient as well;
if the addition of LANS-$\alpha$ were expensive, it would be more
sensible to drop the turbulence model altogether and simply run at
higher resolution.  Thus the choices we make in designing the
LANS-$\alpha$ algorithm and smoothing filters are governed by the
trade-offs of higher-resolution effects versus the cost of adding
LANS-$\alpha$ to the model.  We win this game if we can get the
effects of, say, a doubling of resolution while only increasing the
computation time by a small fraction of the time required to double
the resolution with the standard model.

We have designed a LANS-$\alpha$ algorithm and smoothing filter for
primitive-equation ocean models that meets this criteria of a
successful turbulence model.  The algorithm is presented by Hecht,
Holm, Petersen, and Wingate in a closely related work
\cite{Hecht_ea07jcp}.  The focus of the present paper is the method
used to smooth the advecting velocity.  In the typical LANS-$\alpha$
model derivation, the smooth velocity, $\bu$, is computed from the
rough velocity, $\bv$, with an inverse Helmholtz operator, \bea \bu =
\left( 1-\alpha^2 \nabla^2\right)^{-1}\bv \eea where the namesake
$\alpha$ parameter is a length scale that determines the amount of
smoothing \cite{Holm99pd}.  The Helmholtz inversion is the basis of comparison for
other smoothing methods in this work, as this is the relationship
between $\bu$ and $\bv$ in the original LANS-$\alpha$ equations.  With
it we are able to run successful simulations of LANS-$\alpha$;
however, it requires an iterative solver and is therefore relatively
expensive.  Fortunately, local averaging methods work well in
LANS-$\alpha$ and are much cheaper.  In these methods, the smooth
velocity is computed as a weighted average of nearby neighbors, and can be written as a convolution with the filter function $g$ as
\bea
\bu = g \ast \bv = \int g(\bx-{\bf y})\bv d^3y.
\label{convolution}
\eea
The LANS-$\alpha$ equation using a filter (sometimes called the Kelvin-filtered Navier-Stokes equation) retains Kelvin's circulation theorem \cite{Foias_ea01pd}, which is a property of fundamental importance to LANS-$\alpha$.  If the filter is properly designed, other characteristics of LANS-$\alpha$ with the Helmholtz inversion are guaranteed: $\bu$ is smoother that $\bv$, the energy is well defined, and LANS-$\alpha$ conserves energy in the absence of dissipation.  In this paper we develop design criteria for filters that satisfy these requirements.

There is a precedent of using simple filters in LANS-$\alpha$ in large
eddy simulation (LES) models.  Geurts and Holm \cite{Geurts_Holm05jt}
used a simple three-point top-hat filter as a smoothing operator in
LANS-$\alpha$ and Leray LES modeling of three-dimensional isotropic
turbulent mixing.  They found that the LANS-$\alpha$ and Leray models
are considerably more accurate than dynamic eddy-viscosity models, and
at a lower computational cost.  The LANS-$\alpha$ and Leray models
performed particularly well at capturing the flow features
characteristic of the smaller resolved scales.

This paper is organized as follows: \S\ref{s_LANSa} briefly reviews
the LANS-$\alpha$ equations for primitive-equations and the
POP-$\alpha$ algorithm.  The pressure-velocity instability, which
develops with the incorrect choice of filter weights, is discussed in
\S\ref{s_instability}.  Using an eigenvalue analysis of the
discretized filter in \S\ref{s_evals}, we develop design criteria for
the filter weights to ensure the proper smoothing and kinetic energy
properties.  The results section, \S\ref{s_results}, shows that
POP-$\alpha$ can produce results qualitatively similar to those of
higher resolution simulations of standard POP using either the
Helmholtz inversion or an filter to smooth the velocity.  In the final
section, \S\ref{s_conclusions}, we compare the merits of each.  The
family of filters presented here produce stable simulations and are
20---50\% faster than Helmholtz inversions.  Filters produce better
temperature statistics in the baroclinic instability model problem,
while the Helmholtz inversion produces higher kinetic and eddy kinetic
energy.

\section{The LANS-$\alpha$ model \label{s_LANSa}}
This section gives a brief overview of the primitive-equation version of LANS-$\alpha$ that is used in the POP-$\alpha$ code.  These equations were originally derived by Holm et al. \cite{Holm_ea98inbk}, and are described at length, along with the POP-$\alpha$ algorithm, by Hecht et al. \cite{Hecht_ea07jcp}.  The model equations are
\bea
&\dd{\bv}{t} + \bu\cdot\del\bv + u_3\p_z\bv
   + v_j \del u_j 
   + {\bf f}\times \bu 
 = -\frac{1}{\rho_0}\del \pi 
   + {\cal F}\left({\bv}\right),
\label{NS}\\
& \dd{\pi}{z}=-\rho g,\label{hydrostatic} \\
&\dd{\varphi}{t} + \bu\cdot\del\varphi 
 = {\cal D}\left({\varphi}\right),
\label{tracer} \\
& \del\cdot\bu + \p_z u_3 = 0\label{cont}, \\
& \bu = \left(1-\alpha^2\del^2\right)^{-1}\bv,
\label{Helm} \\
&\pi = p -\frac{1}{2}|\bu|^2 - \frac{\alpha^2}{2}|\del \bu|^2
\label{pi}
\eea where $(v_1,v_2,v_3)=(\bv,v_3)$ and $(u_1,u_2,u_3)=(\bu,u_3)$ are
the rough and smooth velocities, $\del$ is the horizontal gradient,
$\varphi$ is a tracer (temperature, salinity, or a passive tracer),
$\rho$ is the full density, $\rho_0$ is the background density, $\pi$
is a modified pressure, $p$ is the physical pressure, ${\cal F}$ and
${\cal D}$ are momentum and tracer diffusion terms \cite{POP_manual},
${\bf f}$ is the Coriolis parameter, $g$ is gravitational
acceleration, and $\alpha$ is the alpha model's smoothing length
scale.  These equations are hydrostatic (\ref{hydrostatic}),
incompressible (\ref{cont}), and Boussinesq ($\rho_0$ appears in eqn
\ref{NS} but not in \ref{hydrostatic}).

The principle feature of the LANS-$\alpha$ model is that the momentum
equation (\ref{NS}) is an advection-diffusion equation for a
Lagrangian-averaged velocity $\bv$, while the advecting velocity is an
Eulerian-averaged velocity $\bu$.  These names arise in the derivation
of the LANS-$\alpha$ model \cite{Holm99pd}, where velocities are
averaged along a particle track (Lagrangian) or at a particular
location (Eulerian).  The Helmholtz relation (\ref{Helm}) indicates
that $\bu$ is smoother that $\bv$. In this paper we prefer the names
{\it rough velocity} for $\bv$, which is computed prognostically from
(\ref{NS}), and {\it smooth velocity} for $\bu$, which is computed
diagnostically from (\ref{Helm}).  The LANS-$\alpha$ model has an
additional nonlinear term $\sum_{j=1}^3 v_j \del u_j$ in the momentum
equation (\ref{NS}), which does not appear in the Leray
model\cite{Holm99pd}, and which allows LANS-$\alpha$ to satisfy
Kelvin's circulation theorem while conserving energy and a form of
potential vorticity in the absence of dissipation \cite{Holm99pd}.

In this paper, we investigate how to most effectively and efficiently
smooth the rough velocity $\bv$ to obtain the smooth velocity $\bu$.
This could be with a Helmholtz inversion, as in (\ref{Helm}), or with
a convolution using various filter functions, as in
(\ref{convolution}).  LANS-$\alpha$ using a filter rather than a
Helmholtz inversion still satisfies Kelvin's circulation theorem
\cite{Foias_ea01pd}.  If the proper filters are chosen, one also
conserves energy and a form of potential vorticity in the absence of
dissipation.

The most commonly used form of the LANS-$\alpha$ equations use a
three-dimensional (3D) smoothing. In the primitive equations, which
assume a thin layer approximation, this reduces to a 2D smoothing, and
so our filters are horizontal-only as well.  3D smoothing may be tested at
a future time, but we expect that the computational cost would be too
high to justify the benefit.  The Helmholtz inversion is horizontally
isotropic.  Thus one of the design criteria for the smoothing filter
is that it is horizontally isotropic as well.

The POP primitive-equation ocean-climate model
\cite{Smith_ea92pd,Dukowicz_Smith94jgro}, developed and maintained at
Los Alamos National Laboratory, was augmented to include the
LANS-$\alpha$ formulation \cite{Hecht_ea07jcp}.  POP is a split
implicit/explicit code; that is, it takes an implicit step for the
barotropic (vertically integrated 2D) velocity field, and takes an
explicit step for the baroclinic (remaining 3D) velocity field.  This
is done so that the timestep is not limited by the fast free-surface
gravity waves in the barotropic velocity field.

In the LANS-$\alpha$ version of POP, named POP-$\alpha$, a smooth
velocity field must be computed for both the baroclinic and barotropic
velocity fields.  In the {\it explicit} baroclinic component of the
code, the smooth baroclinic velocity, $\bu$, is computed by simply
smoothing $\bv$.  The {\it implicit} barotropic component is more
complicated, and in the end a reduced version of the full barotropic
algorithm, as derived from the LANS-$\alpha$ equations, was used
\cite{Hecht_ea07jcp}.  The reduced algorithm produced nearly identical
results to the full algorithm, but was three to four times faster.

A few pieces of the POP-$\alpha$ barotropic algorithm are needed to understand the pressure-velocity instability presented in the following section.  The 2D barotropic velocities, denoted by capital letters, are found by integrating the full velocity from the bottom to the free surface height, $\eta$:
\bea
\bU \equiv \frac{1}{H+\eta}\int_{-H}^\eta \bu \;\;dz,
\eea
where $H(x,y)$ is the ocean depth when the surface is at rest.  By integrating the continuity equation (\ref{cont}), which states that the {\it smooth} velocity is divergence-free, we obtain a prognostic equation for the free surface height,
\bea
& \int_{-H}^\eta \left( \del\cdot\bu + \p_z u_3 \right) dz=0,\\
& \dd{\eta}{t} + \del \cdot \left(H+\eta\right)\bU = 0.\label{eta}
\eea
The free surface height now replaces the pressure in the barotropic form of the momentum equation,
\bea
&\ds \dd{\bU}{t} +  {\bf f}\times\bU
  = - g\del\eta +  {\bf G}, \label{barotropic_NS}
\eea
where ${\bf G}$ contains the vertically integrated forcing terms.  The barotropic momentum equation is the same as the shallow water equation.

\section{Pressure-velocity instability \label{s_instability}}

In this section, we present an analytic stability analysis of a
pressure-velocity instability that constrains the filter weights in
the barotropic filter.  The pressure-velocity instability is a
phenomenon particular to models where a smoothed velocity field is
used in the equation for free surface height $\eta$ (\ref{eta}).
Consider a velocity field in one-dimensional (1D) discretized shallow
water equations, where the pressure and velocity gridpoints are
offset.  If the velocity field were at the Nyquist frequency
(alternating between positive and negative at each grid cell), then
using the standard shallow water equations, the free surface height
would raise (lower) where the velocity field converges (diverges).  In
either case, a pressure gradient force would ensue that would work
against the tendency of the surface to undergo further deformation,
providing a stabilizing effect.

\begin{figure}[tbh]
\center
\begin{tabular}[c]{lcc}
\scalebox{.8}{\includegraphics{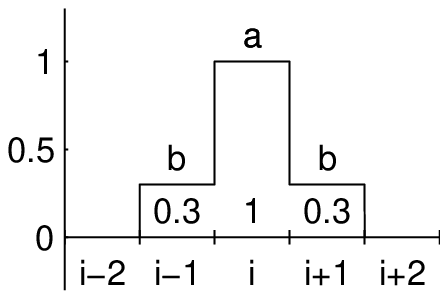}}  &
\scalebox{.7}{\includegraphics{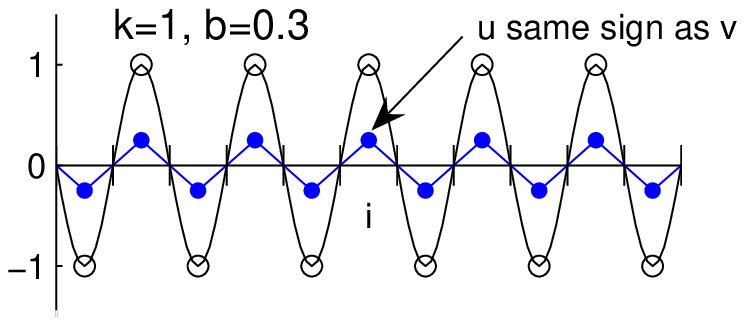}}  &
\scalebox{.7}{\includegraphics{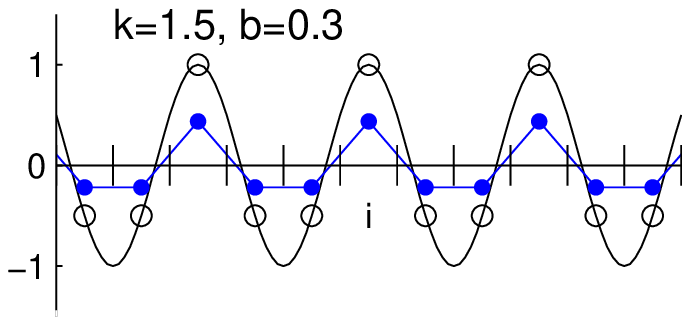}} \\
\scalebox{.8}{\includegraphics{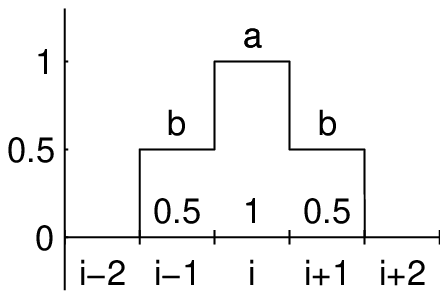}}  &
\scalebox{.7}{\includegraphics{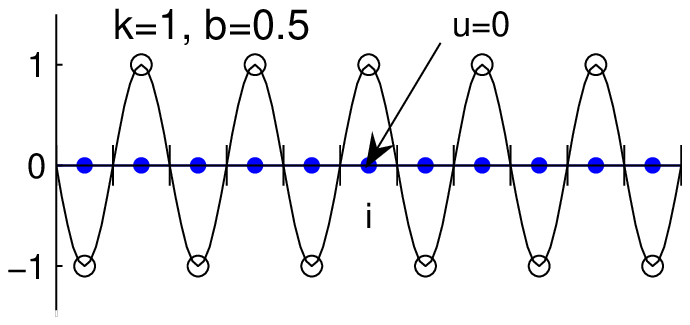}}  &
\scalebox{.7}{\includegraphics{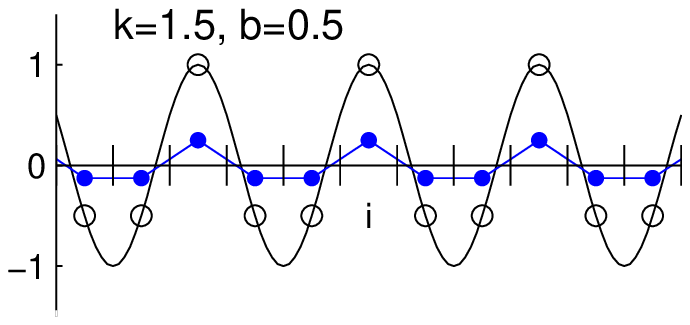}} \\
\scalebox{.8}{\includegraphics{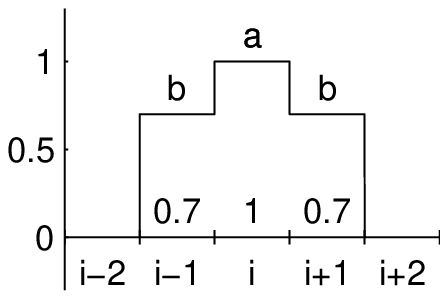}}  &
\scalebox{.7}{\includegraphics{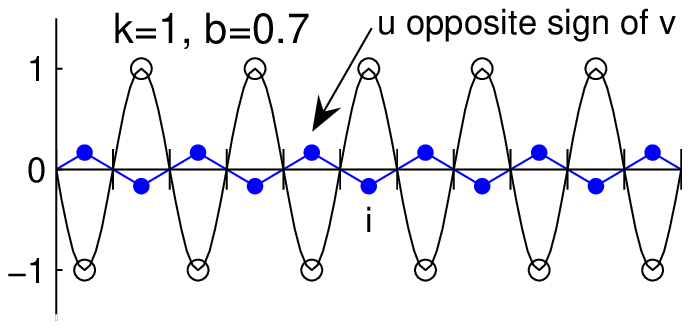}}  &
\scalebox{.7}{\includegraphics{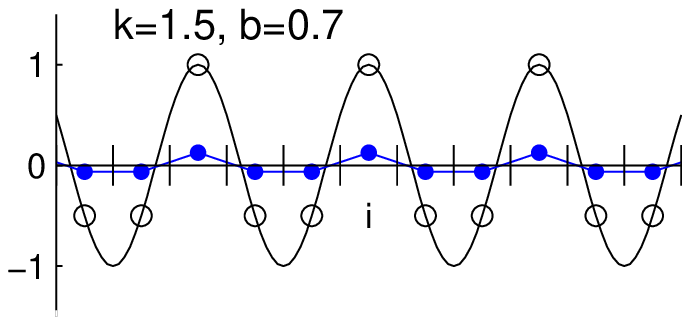}} 
\end{tabular}
\caption{\label{f_F3} 
The effects of filtering the rough velocity $v$ (open circles) to compute the smooth velocity $u$ (closed circles) for a 1D filter of width three, where the weight of the neighboring cell, $b$, is varied between 0.3 and 0.7, as shown in the left column.  For the Nyquist frequency ($k=1$, middle column) $u$ is of opposite sign to $v$ if $b>0.5$; this results in an instability, since the equation for free surface height $\eta$ involves $u$.  For a slightly higher wavenumber ($k=1.5$, right column) $u$ has the same sign as $v$ for $b<1$.
}\end{figure}

In the LANS-$\alpha$ model the equation for the free surface height involves the {\it smooth} velocity because this equation is derived from the continuity equation (\ref{cont}), which involves the smooth velocity.  When the outer stencil weights are sufficiently large in LANS-$\alpha$, a {\it pressure-velocity instability} ensues in the highest frequencies.  This can be shown with the simplest case, a 1D velocity field and a smoothing stencil of width three,
\bea
u_i = \frac{bv_{i-1}+av_i+bv_{i+1}}{a+2b}. \label{F3}
\eea
In most cases, $u_i$ and $v_i$ will have the same sign at each gridpoint, as one would expect for a smoothing operator.  However, for high frequencies and sufficiently large outer stencil weight $b$, $u$ can have the opposite sign of $v$, as shown in Fig. \ref{f_F3} for the Nyquist frequency and $b=0.7$.  

\begin{figure}[tbh]
\center
\begin{tabular}[c]{cc}
\scalebox{1.}{\includegraphics{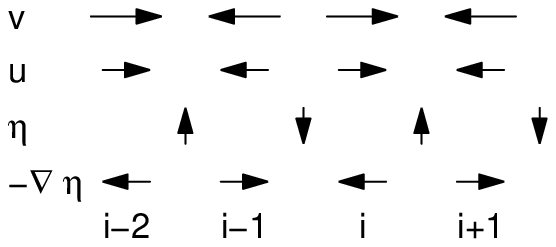}}  &
\scalebox{1.}{\includegraphics{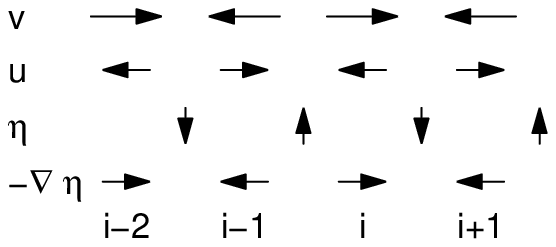}} \\
(a) stable case, $b<0.5$ & 
(b) unstable case, $b>0.5$ 
\end{tabular}
\caption{\label{f_1D_diagram} 
Schematic of variables to illustrate the pressure-velocity feedback instability.  In the stable case (a) the smooth and rough velocities, $u$ and $v$, are of the same sign, and the pressure gradient force $-\del\eta$ counters the original velocity perturbation, as expected.  In the unstable case (b), $u$ and $v$ have opposite signs;  the free surface height, which is computed from $u$, increases (decreases) where $u$ converges (diverges);  now the pressure gradient force reinforces the rough velocity $v$, increasing this perturbation.
}\end{figure}

The tendency equation for $v$ depends on $\eta$, but $\eta$ in turn is
not directly dependent on the convergence/divergence of $v$ but on that
of $u$. In the case of a pure harmonic mode, restricting the local filter
to produce a smoothed velocity $u$ of the same sign everywhere as the
rough velocity $v$ is sufficient to also ensure that the divergence of
smooth and rough velocities share the same sign, preserving the essential
physical feedback of the underlying equations even after inclusion of the
LANS-$\alpha$ model. Cases in which this physical feedback is respected
or not, depending on the filter width, are illustrated in Fig.
\ref{f_1D_diagram}.

This understanding of the pressure-velocity instability allows us to construct an analytical stability criterion as follows: {\it The filter weights must be chosen so that the filtered velocity $u$ retains the same sign as $v$ for all harmonic modes resolved on the grid.}  

For example, for the stencil of width three (\ref{F3}) where $a$ is normalized to 1, consider the Nyquist frequency
\bea
v_{i+j} = \cos(\pi j/k)
\eea
with $k=1$.
Since $v_i=1$, the condition that $u$ and $v$ are like-signed is $u_i>0$.  Using (\ref{F3}), the constraint on the stencil is $b<0.5$.  For a slightly higher wavenumber, $k=1.5$, the constraint is weaker, $b<1$.  For still higher wavenumbers, $u$ and $v$ are like-signed for any choice of $b$.  Thus $b$ must be chosen based on the tightest constraint, $b<0.5$.  

\begin{figure}[tbh]
\center
\begin{tabular}[c]{cc}
\scalebox{.5}{\includegraphics{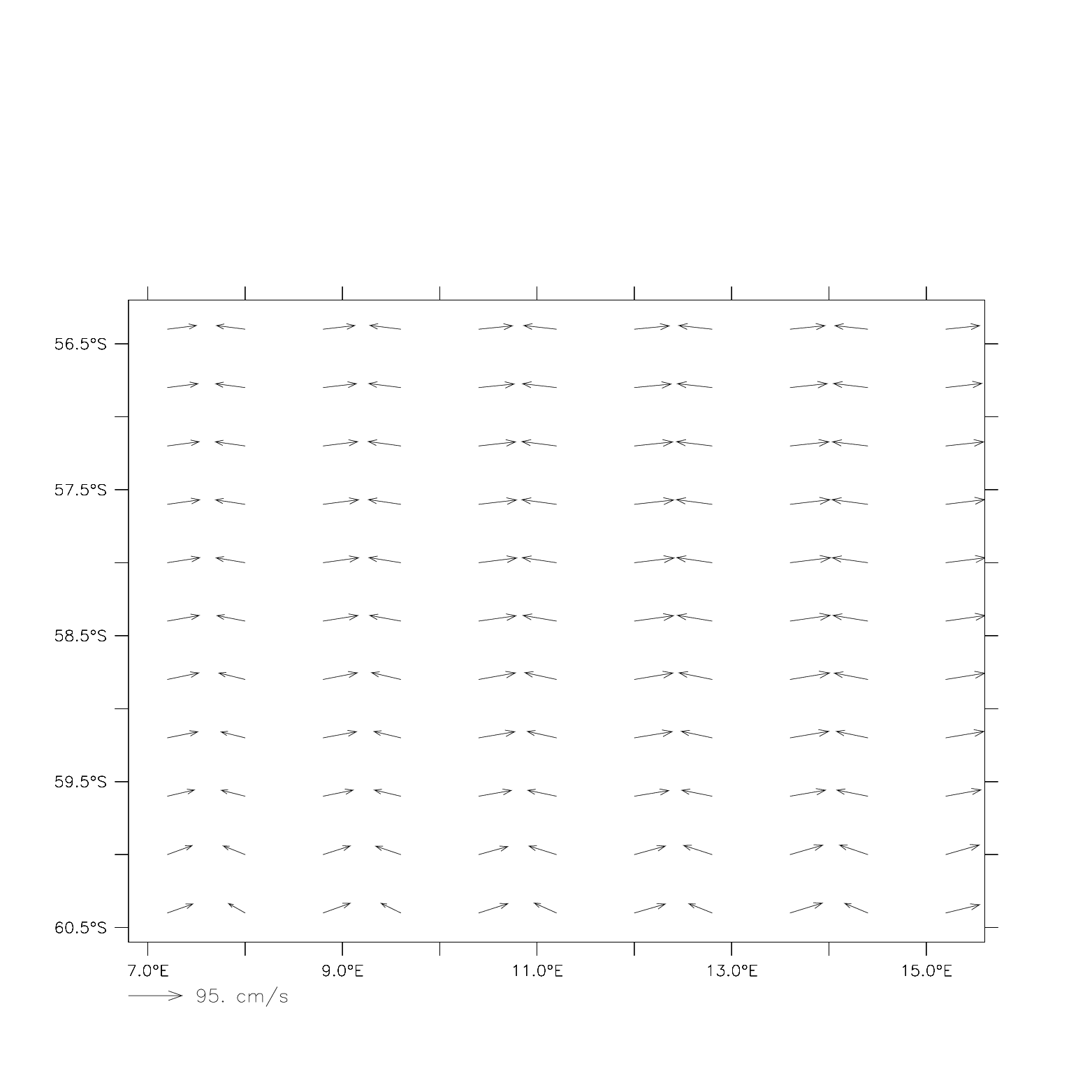}} &
\scalebox{.5}{\includegraphics{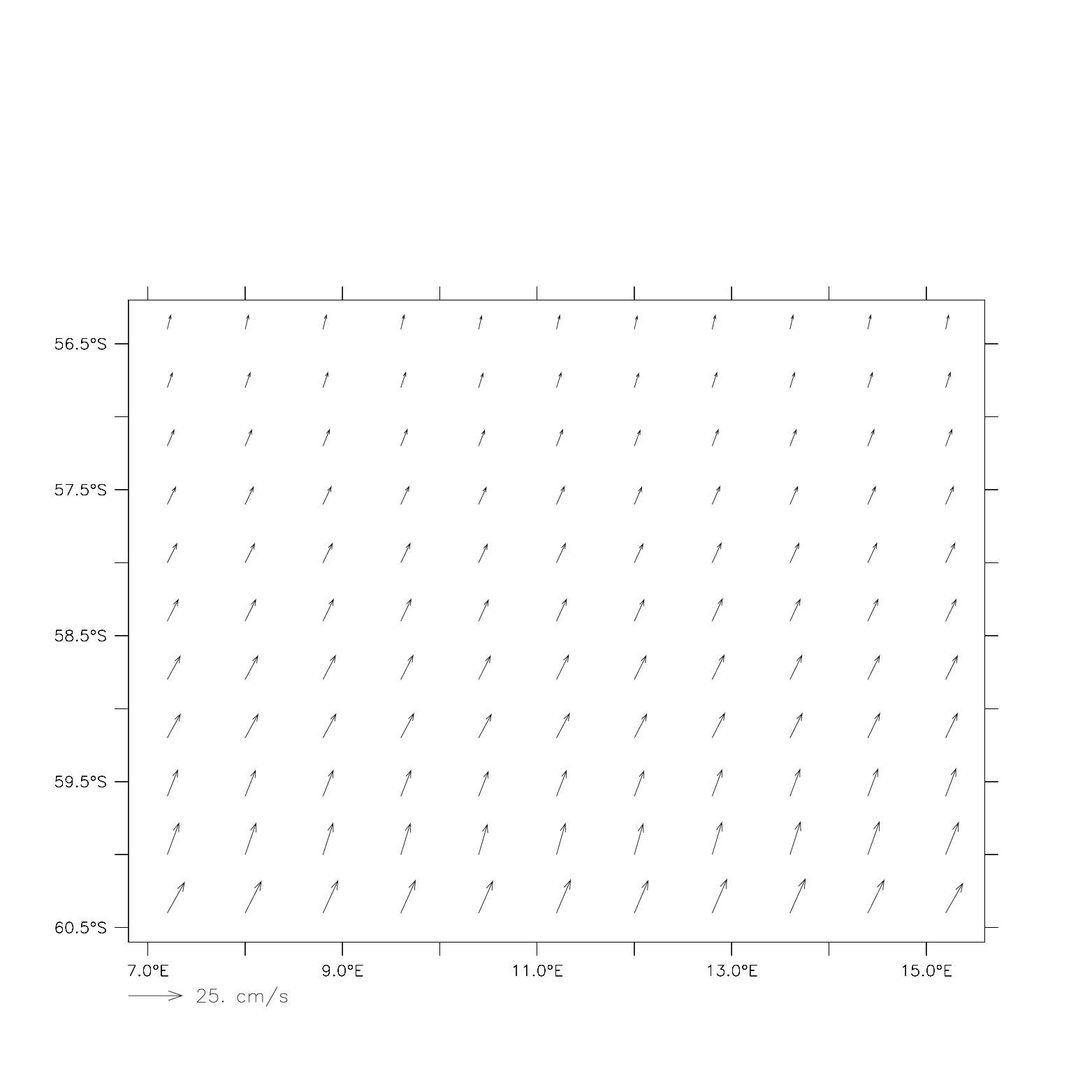}} \\
(a) rough velocity $v$ & 
(b) smooth velcotiy $u$
\end{tabular}
\caption{\label{f_vel_ex} 
An example of the pressure-velocity instability.  Shown are the velocity fields of simulation 0.8F3 using $b=0.52$ after 100 time steps.  The Nyquist frequency grows exponentially in the rough velocity $v$ (a), but is filtered out of the smooth velocity $u$ (b) (note difference in vector length scales).  The surface elevation $\eta$ is computed using $u$, so pressure gradient forces don't counteract this unstable mode.
}\end{figure}

Numerical simulations of POP-$\alpha$ with a stencil of width three almost exactly correspond to the predictions of this stability analysis.  The model runs stably when $b<0.48$.  At larger values of $b$ an instability quickly grows in $\bv$ and $\eta$ at the Nyquist frequency, so that $\bv$ is four times larger than $\bu$ within 100 time steps (Fig \ref{f_vel_ex}).  By 150 timesteps the conjugate gradient routine in the barotropic solver fails to converge within 1000 iterations.  Numerical simulations use a 2D stencil that is a squared version of the 1D stencil, as discussed below.

\begin{landscape}

\begin{figure}[tbh]
\center
\begin{tabular}[c]{lccccc}
& 
\scalebox{.7}{\includegraphics{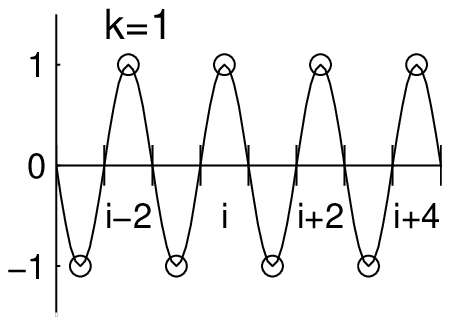}}  &
\scalebox{.7}{\includegraphics{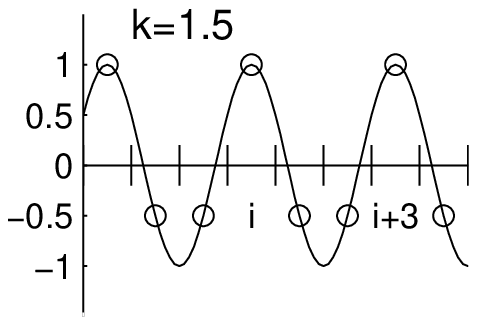}}  &
\scalebox{.7}{\includegraphics{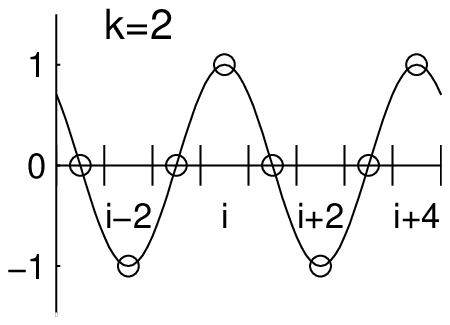}}  &
\scalebox{.7}{\includegraphics{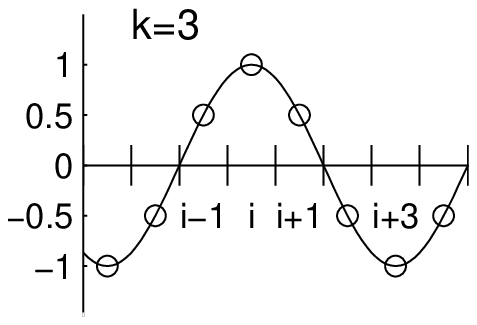}}  &
\scalebox{.7}{\includegraphics{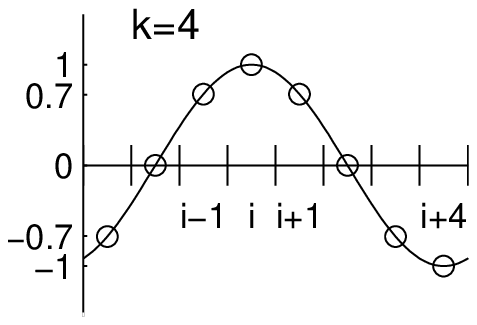}}  \\
condition:&$1+2b(-1)$ &
$1+2b\left(-\frac{1}{2}\right)$ & 
$1+2b(0)+2c(-1)$ &
$1+2b\left(\frac{1}{2}\right)$ & 
$1+2b\left(\frac{\sqrt{2}}{2}\right)$ 
\\
&$+2c(1)+2d(-1)$ &
$+2c\left(-\frac{1}{2}\right) + 2d(1)$ & 
$+2c(-1)+2d(0)$ &
$+2c\left(-\frac{1}{2}\right)+2d(-1)$ & 
$+2c(0)+2d\left(-\frac{\sqrt{2}}{2}\right) $ 

\\
&$\;\;+2e(1)>0$&
$\;\;+2e\left(-\frac{1}{2}\right)>0$&
$\;\;+2e(-1)>0$&
$\;\;+2e\left(-\frac{1}{2}\right)>0$&
$\;\;+2e(-1)>0$

\\
\scalebox{.7}{\includegraphics{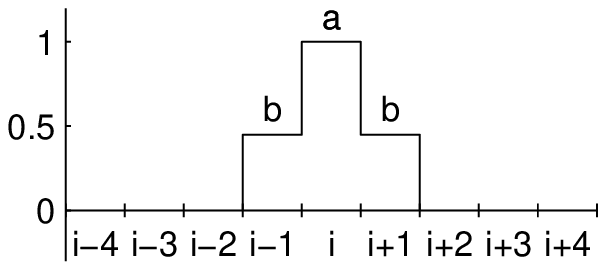}}  &
\fbox{$b<\frac{1}{2}$} &$b<1$ & stable $\forall\:\: b$ & stable $\forall\:\: b$ & 
stable $\forall\:\: b$ 
\\
\scalebox{.7}{\includegraphics{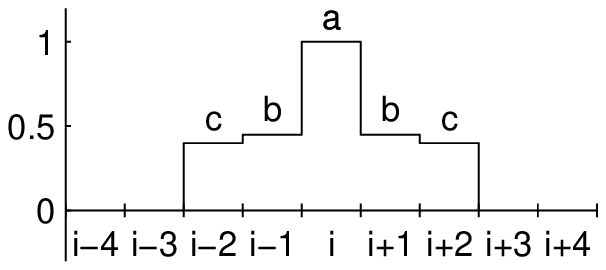}}  &
$b<\frac{1}{2}+c$ & \fbox{$b+c<1$} & \fbox{$c<\frac{1}{2}$} &
  $c<1+b$& stable 
$\forall\:\: b,c$ 
\\
\scalebox{.7}{\includegraphics{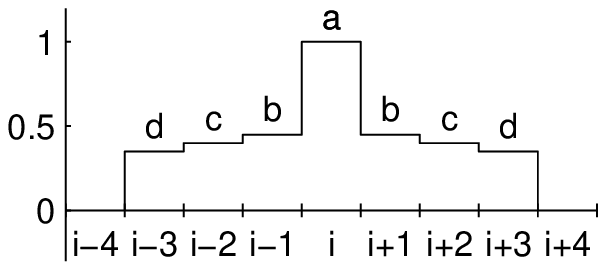}}  &
\fbox{$b+d<\frac{1}{2}+c$} &$b+c<1+2d$ & \fbox{$c<\frac{1}{2}$} & 
\fbox{$c+2d<1+b$} & 
$d<\frac{\sqrt{2}}{2}+b$ 
\\
\scalebox{.7}{\includegraphics{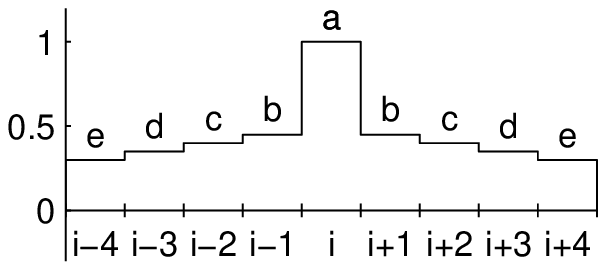}}  &
$b+d<\frac{1}{2}+c+e$ &\fbox{$b+c+e<1+2d$} & 
$c<\frac{1}{2}+e$ & \fbox{$c+2d+e<1+b$} & 
\fbox{$\sqrt{2}d+2e<1+\sqrt{2}b$} 

\end{tabular}
\caption{\label{f_filters} 
Conditions required for stability for filters of width 3, 5, 7, and 9 (rows) for several wavenumbers of $v$ (columns).  Variables $a$---$e$ are the weights of the 1D stencil, normalized such that $a=1$.  A stable scheme requires that $u_i$, the smooth velocity, is the same sign as $v_i$.  Conditions that limit the weights $b$---$e$ to be less that one-half are boxed.
}\end{figure}

\end{landscape}

\begin{figure}[tbh]
\center
\begin{tabular}[c]{cc}
\scalebox{.8}{\includegraphics{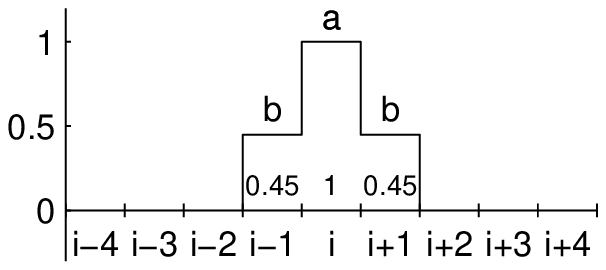}} &
\scalebox{.8}{\includegraphics{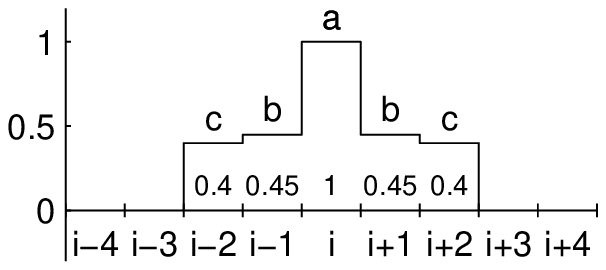}} \\
(a) 1D stencil, width 3 & (b) width 5 \\
\scalebox{.8}{\includegraphics{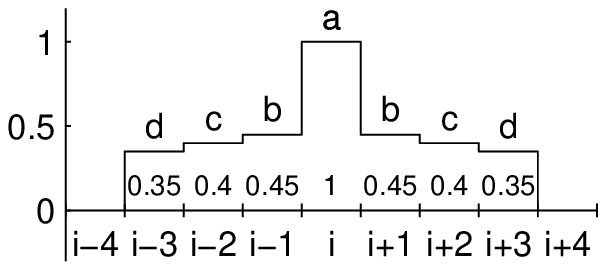}} &
\scalebox{.8}{\includegraphics{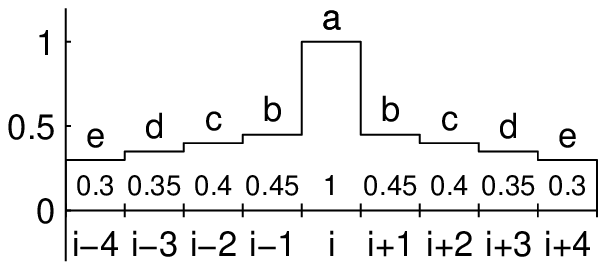}} \\
(c) width 7 & (d) width 9
\end{tabular}
\caption{\label{f_filters_1D} 
Filters used in this study, with the value of each weight $b$---$e$ under the corresponding variable.
}\end{figure}

Because of the limitation imposed by the pressure-velocity
instability, further smoothing must be accomplished by increasing the
stencil size.  The criterion that $u$ and $v$ must be like-signed for
all frequencies applies, but larger stencils have more unknowns, one
for each stencil weight.  Figure \ref{f_filters} shows these
constraints in chart form for 1D filters of width 3, 5, 7, and 9 and
for wavenumbers $k=1\ldots4$.  In general, these constraints require
that the weights are each less than $0.5$.  Beyond that, each
consecutive weight moving away from the center must be slightly less
than the one before it.  This latter requirement can be seen in the
constraints for the stencil of width nine for wavenumber $k=3$ and
$k=4$, 
\bea
&c+2d+e<1+b,\\
&\sqrt{2}d+2e<1+\sqrt{2}b.  
\eea 
Neither of these are satisfied if all
the weights are $0.5$, but are satisfied when the weights decrease as
\bea
b=0.45,\;\;c=0.4,\;\;d=0.35,\;\;e=0.3.  \label{weights} 
\eea
Again, the results of this stability analysis were verified using the
POP-$\alpha$ model.  Simulations using the stencil weights in
(\ref{weights}) were always stable (Fig \ref{f_filters_1D}).  If any
of the weights are increased by 0.02, a fast-growing, high frequency
appears in $\bv$ and $\eta$ (Fig \ref{f_vel_ex}).

\begin{figure}[tbh]
\center  
\begin{tabular}[c]{c}
(a)\scalebox{.8}{\includegraphics{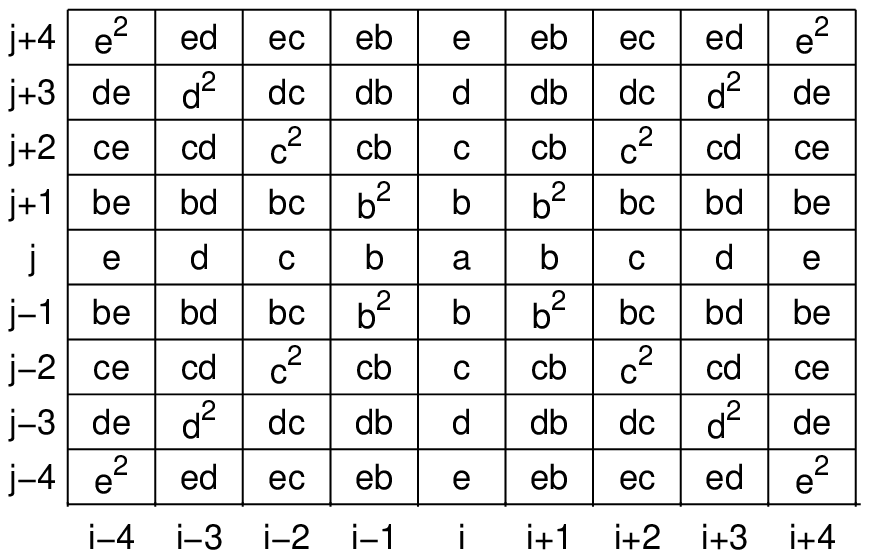}} 
(b)\scalebox{.8}{\includegraphics{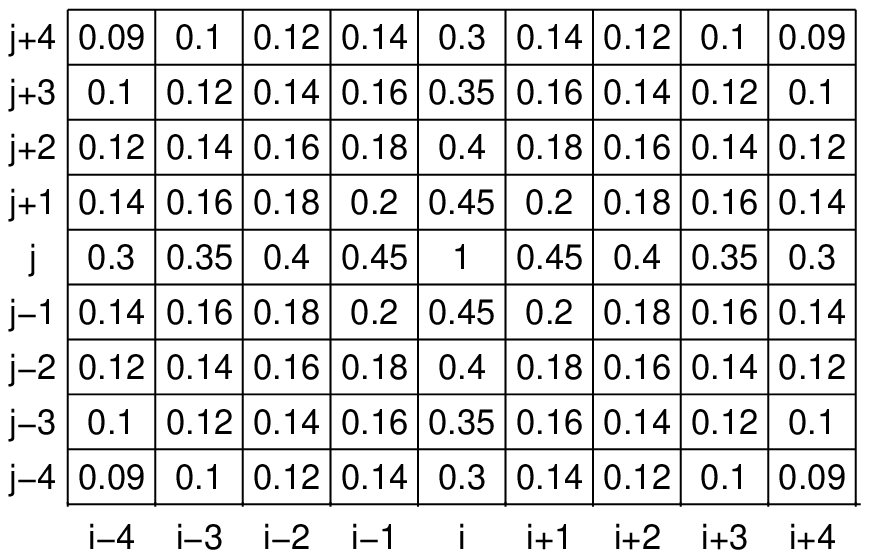}} 
\end{tabular}
\caption{\label{f_filters_2D} 
The 2D version of a filter stencil of width nine general form (a).  This is simply the square of the 1D stencil.  The 2D filters of width 3, 5, and 7 are subsets of this stencil.  Our specific choice of weights is shown in (b).
}\end{figure}

In the POP-$\alpha$ model the smoothing operator is 2D.  The 2D stencil was chosen to be a squared version of the 1D stencil (Fig. \ref{f_filters_2D}).  Such a stencil is isotropic, and thus follows the same stability analysis as the 1D case presented above.  POP-$\alpha$ is very sensitive to these stability constraints; for example, if the corner weights of the stencil in Fig. \ref{f_filters_2D}b are increased by 10---20\%, the simulation is unstable.

Other experiments were conducted where the 2D stencil was a diamond or circle instead of a square.  The motivation here is that in the standard square 2D stencil (Fig. \ref{f_filters_2D}) the corner weights are small, and don't contribute much to the average.  Thus computation time could be saved by removing the corners from the stencil, and the degree of smoothing should be nearly the same.  However, in practice any stencil other than a full square stencil resulted in unstable simulations that did not converge in the barotropic solver.  This was even true when only a single gridpoint was left off of the stencil on each corner, for example when the 0.09 weight is removed from a nine-wide stencil in Fig. \ref{f_filters_2D}b.  Thus all of the simulations presented in this work use a complete square stencil.

Near boundaries, the stencil is reduced in size so that all grid-cells
used in the averaging calculation are water-cells, and none are
land-cells.  For example, at a grid-cell directly beside or diagonal
to a land cell no averaging takes place (i.e. $u_{i,j}=v_{i,j}$).  If
the nearest land-cell is two grid-cells away, a filter of width three
is used.  The disadvantage of this method is that the smoothing goes
to zero as one approaches a land-cell.  This stands in contrast to the
LANS-$\alpha$ equations, where the smoothing scale $\alpha$ is
constant throughout the domain.  Other filtering schemes near the
boundary were considered: One could use a constant-size stencil and
assign zero velocity to land-cells, but this strongly damps the smooth
velocity near the boundary.  Another possibility is to use large but
nonsymmetric stencils near the boundary, but this proved to be
unstable.

\section{Eigenvalue analysis of the filter \label{s_evals}}

Most theoretical results for the LANS-$\alpha$ model depend on the two
velocities being related through the Helmholtz operator. We can
replace the Helmholtz operator with filters if two critical properties
are preserved.  These are that (1) the filtered velocity $\bu$ must be
smoother than the unfiltered velocity $\bv$, and (2) that the energy,
\bea
E = \frac{1}{2}\int \bu\cdot \bv \; d^3 x 
\label{energy}
\eea
must be well defined (e.g., non-negative).  When both of these criterion are satisfied, the filter may be used in place of the Helmholtz inversion operator in the LANS-$\alpha$ model.  In this section we will review why the Helmholtz inversion satisfies these criterion, and then proceed to show that the filter described in the previous section satisfies them as well.

The Helmholtz inversion is easily shown to be a smoothing operator using Fourier analysis,
\bea
{\hat \bu}(k,l) = \frac{\hat \bv(k,l)}{1+\alpha^2 k_h^2},
\eea
where ${\hat \bu}$ and ${\hat \bv}$ are the Fourier coefficients, $(k,l)$ are horizontal wavenumbers and $k_h^2=k^2+l^2$.  For scales much larger than $\alpha$ the Fourier coefficients are unaffected, because $1/k_h>>\alpha$ so $\alpha^2 k_h^2<<1$ and ${\hat \bu} \sim {\hat \bv}$.  For scales at or smaller than $\alpha$ the Fourier coefficients are strongly damped because $\alpha^2 k_h^2>1$ so that ${\hat \bu} < {\hat \bv}$.  The energy for the Helmholtz inversion is also well defined, because (\ref{energy}) can be rewritten, using integration by parts, as
\bea
E = \frac{1}{2}\int \left| \bu \right|^2 
        + \alpha^2  \left| \del \bu \right|^2 \; d^3x,
\eea
which is nonnegative.

\begin{figure}[tbh]
\center  
\begin{tabular}[c]{c}
\scalebox{1}{\includegraphics{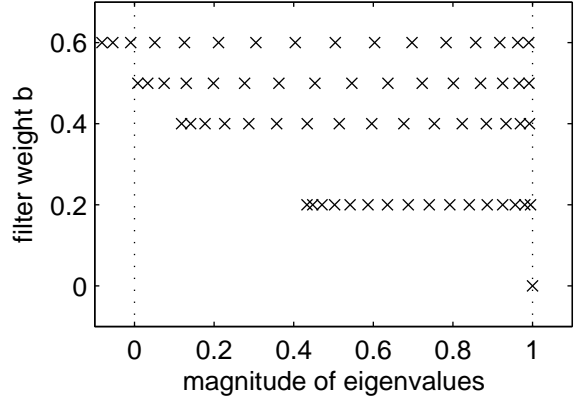}} 
\end{tabular}
\caption{\label{f_eval_1D} 
Eigenvalues of the 1D, width-three filter operator using 16 gridpoints.  When $b=0$ the operator is simply the identity, so all eigenvalues are one.  As $b$ increases the eigenvalues spread out below one, indicating that some eigenvectors contract.  Eigenvalues are never larger than one, so no eigenvectors increase.  When $b>0.5$ there are negative eigenvalues.
}\end{figure}

\begin{figure}[tbh]
\center  
\begin{tabular}[c]{c}
(a)\scalebox{.8}{\includegraphics{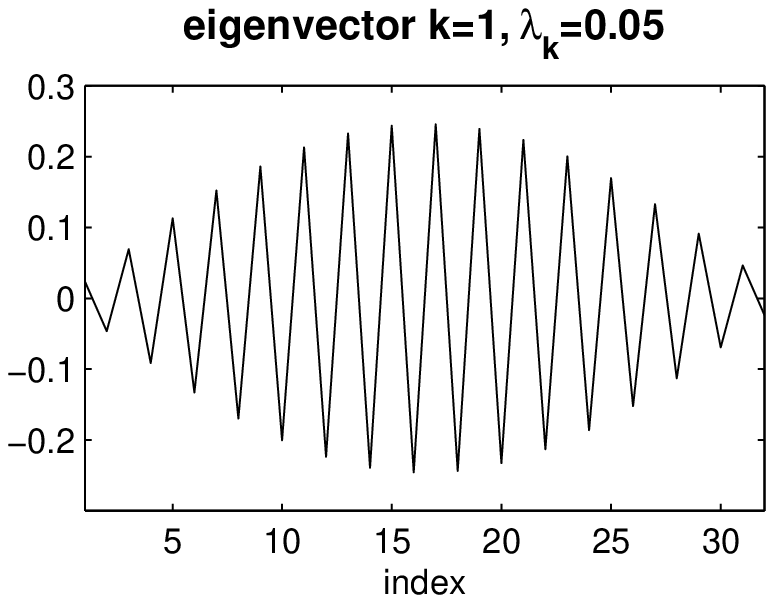}} 
(b)\scalebox{.8}{\includegraphics{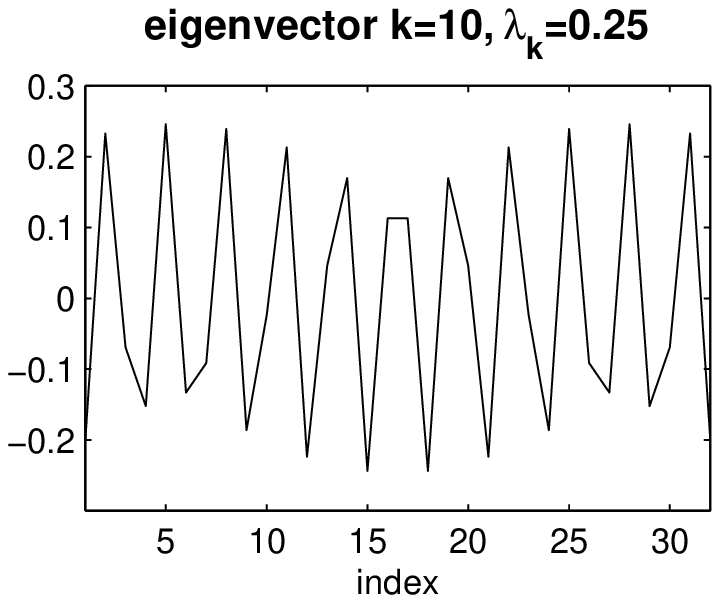}} \\
(c)\scalebox{.8}{\includegraphics{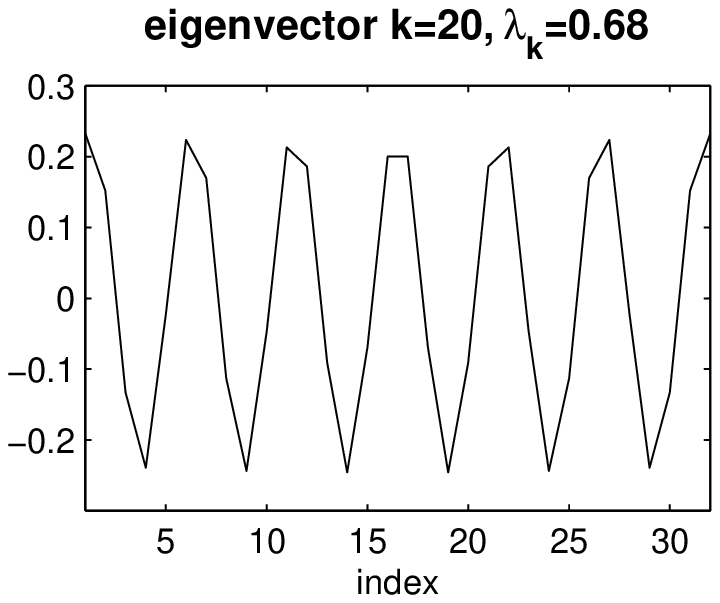}} 
(d)\scalebox{.8}{\includegraphics{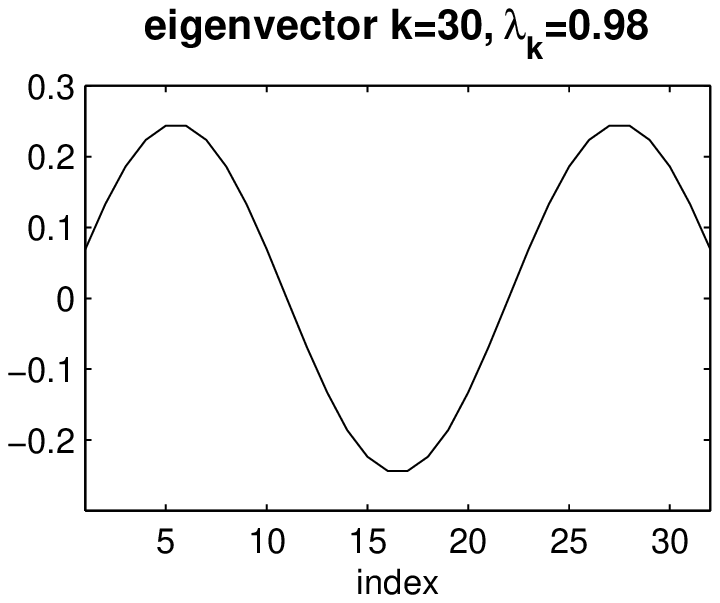}} 
\end{tabular}
\caption{\label{f_evect_1D} 
Eigenvectors of the 1D, width-three filter operator using 32 gridpoints.  High wavenumber modes (a,b) have eigenvalues near zero, while low wavenumber modes (c,d) have eigenvalues near one.  This shows that the filter is a smoothing operator, because the highest wavenumbers are strongly damped.
}\end{figure}

We next show that these properties also hold for our filters by using the eigenvalue problem applied to the discrete filter-operator matrix A.  To illustrate this process, first consider the simple three-wide filter in 1D shown in (\ref{F3}).  This filter can be written as a tridiagonal matrix,
\bea
A = \frac{1}{a+2b}\left[ 
\begin{array}{cccccc}
a & b & 0 & \cdots & 0 \\
b & a & b & 0      & \cdots \\
0 & \ddots & \ddots & \ddots & 0 \\
0 & \cdots & 0 & b & a 
\end{array}
\right].
\eea
so that a vector can be filtered using $\bu=A\bv$.  The eigenvalues of $A$ vary as a function of $b$.  If $b=0$, $A$ is the identity matrix and the eigenvalues are all one.  As $b$ increases, the range of eigenvalues increases but always remains less than one (Fig. \ref{f_eval_1D}).  The smallest eigenvalues are associated with the highest wavenumber modes (Fig. \ref{f_evect_1D}), indicating that the most oscillatory modes are damped the strongest, and that the matrix $A$ is indeed a smoothing operator.

\begin{figure}[tbh]
\center  
\begin{tabular}[c]{c}
\scalebox{1}{\includegraphics{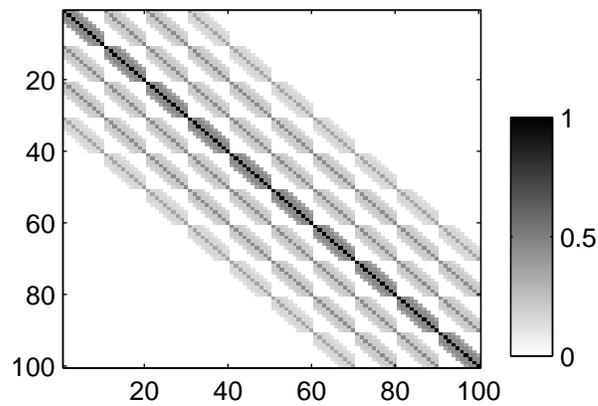}} 
\end{tabular}
\caption{\label{f_matrix_A_2D} 
Matrix representation of the 2D, 7x7 filter using 10x10 gridpoints.  The full matrix is 100 elements square, while each block is 10 square.  This matrix is not yet normalized by the factor $1/(a+2b+2c+2d+2e)^2$, so weights range between 0 and 1.
}\end{figure}

\begin{figure}[tbh]
\center  
\begin{tabular}[c]{c}
\scalebox{1}{\includegraphics{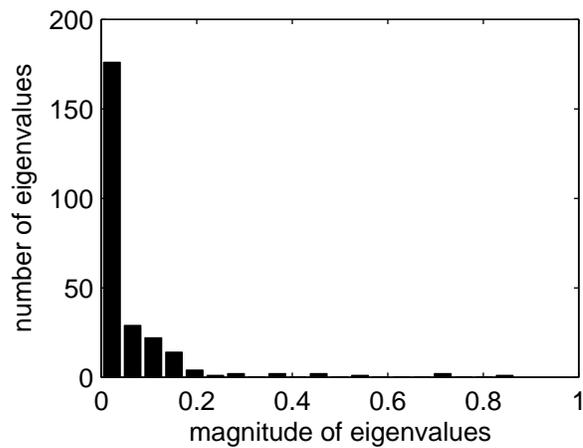}} 
\end{tabular}
\caption{\label{f_evals_2D} 
Distribution of eigenvalues of the 2D, 9x9 filter operator using 16x16 gridpoints.  Filter weights are as shown in Fig. \ref{f_filters_1D}.  The majority of eigenvalues are less than 0.1, indicating that most eigenmodes are stongly damped.
}\end{figure}

\begin{figure}[tbh]
\center  
\begin{tabular}[c]{c}
(a)\scalebox{.8}{\includegraphics{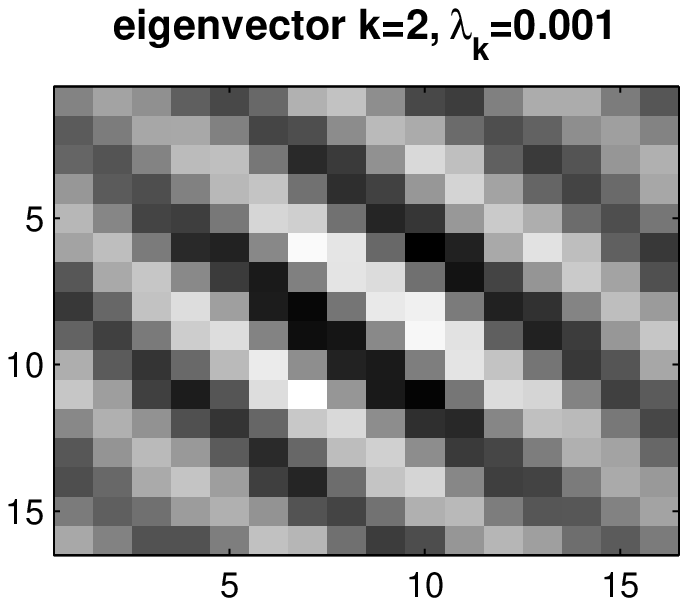}} 
(b)\scalebox{.8}{\includegraphics{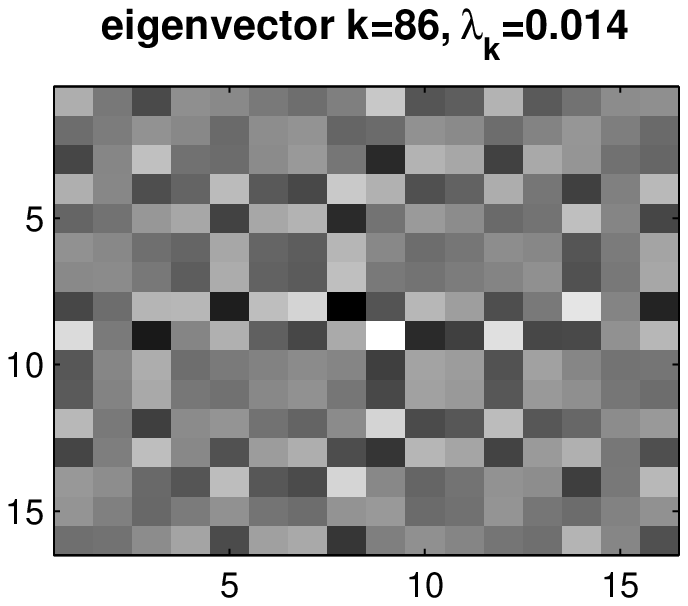}} \\
(c)\scalebox{.8}{\includegraphics{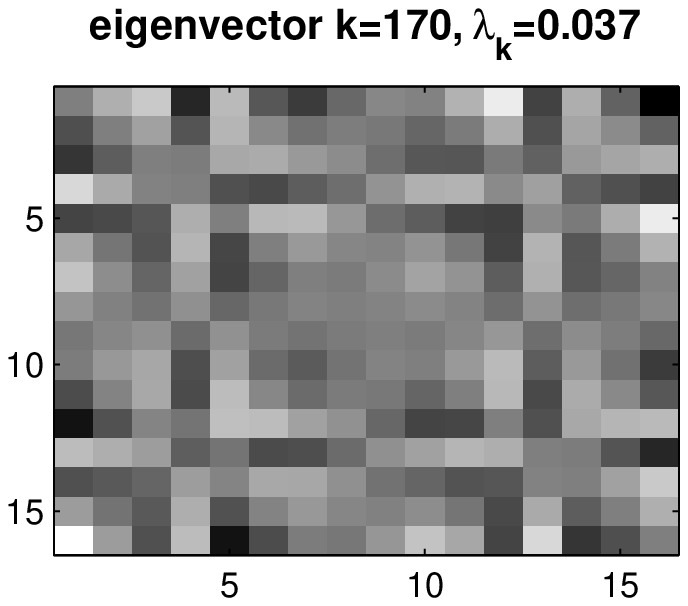}} 
(d)\scalebox{.8}{\includegraphics{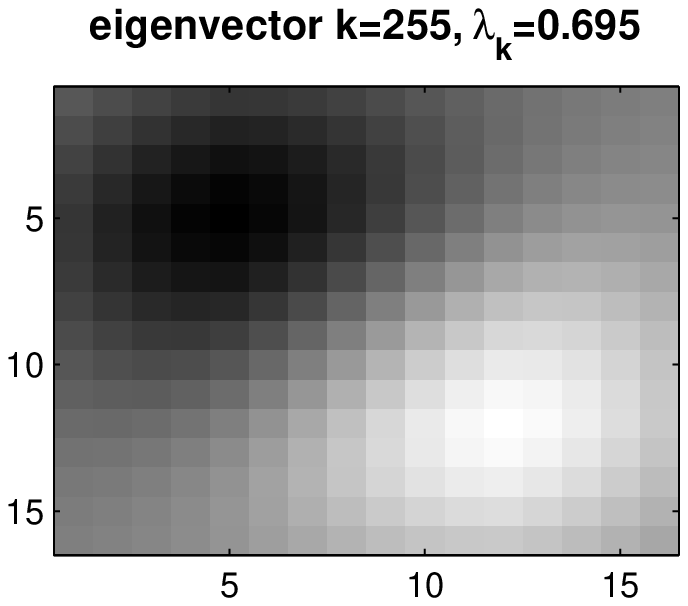}} 
\end{tabular}
\caption{\label{f_evect_2D} 
Eigenmodes of the 2D, 9x9 filter operator using 16x16 gridpoints.  As in the 1D case, highly oscillatory modes (a) have eigenvalues near zero and so are strongly damped, while less oscillatory modes (d) have eigenvalues near one and so are only weakly damped.  This shows that the filter we have implemented is a smoothing operator.
}\end{figure}

We now evaluate the eigenvalues of the full filter (Fig. \ref{f_filters_2D}) in 2D.  On an $n\times n$ grid the filter operator is an $n^2$ matrix (Fig. \ref{f_matrix_A_2D}).  Here we consider Dirichlet boundary conditions of zero, but the results are similar for periodic boundaries.  As in the 1D case, all eigenvalues are less than one.  We have chosen weights so that most eigenvalues are near zero but positive (Fig. \ref{f_evals_2D}).  The most oscillatory modes have very small eigenvalues, so that these are strongly damped (Fig. \ref{f_evect_2D}).  This ensures that the filter is a smoothing operator.

\begin{figure}[tbh]
\center  
\begin{tabular}[c]{c}
(a)\scalebox{.9}{\includegraphics{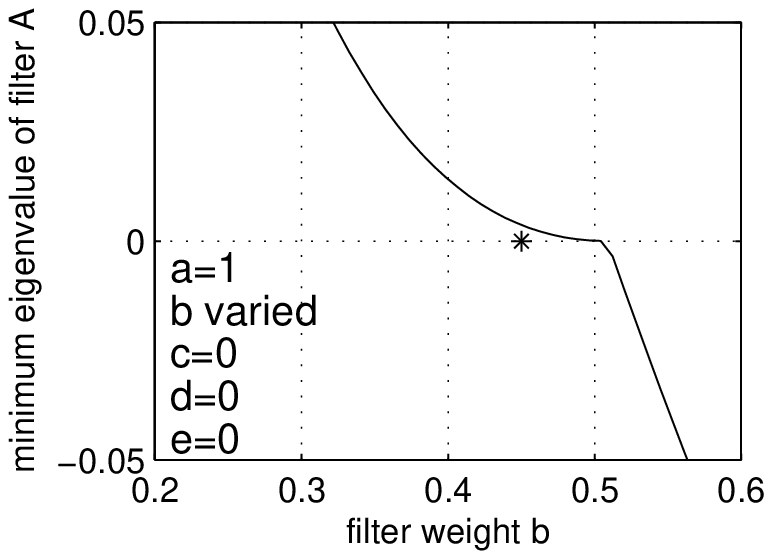}} 
(b)\scalebox{.9}{\includegraphics{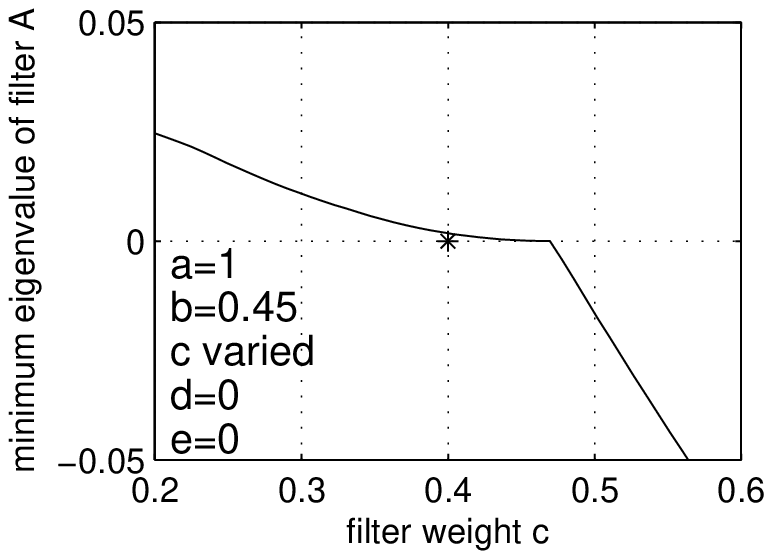}} \\
(c)\scalebox{.9}{\includegraphics{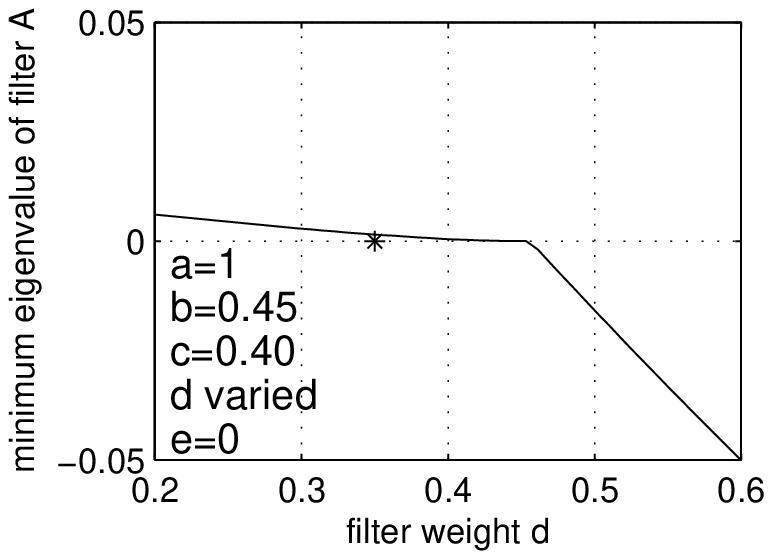}} 
(d)\scalebox{.9}{\includegraphics{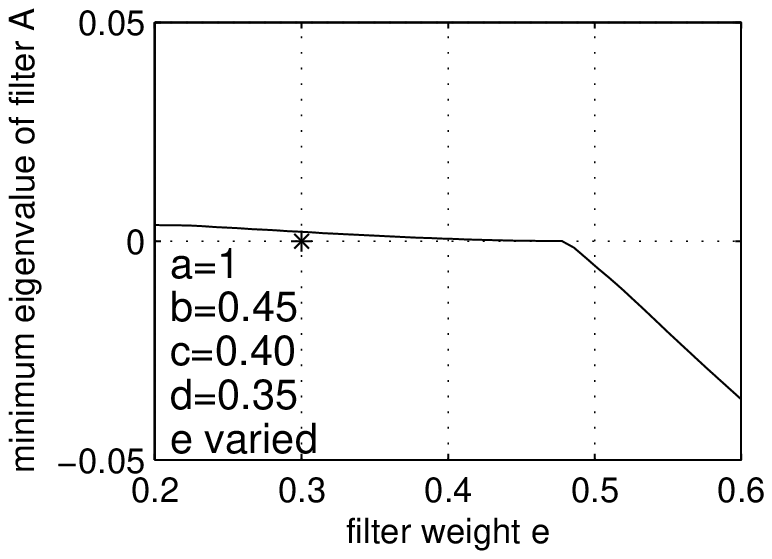}} 
\end{tabular}
\caption{\label{f_pos_evals} 
Minimum eigenvalue of the 2D, 9x9 filter operator using 16x16 gridpoints, as a function of each filter weight.  If all eigenvalues are strictly positive then the operator is positive definite, and the energy is well defined.  The value of each filter weight is chosen (starred points) to satisfy this criterion.
}\end{figure}

The last task is to ensure that the energy defined in (\ref{energy}) is well defined.  The matrix $A$ is positive definite if $\bv^TA\bv>0$ for all nonzero vectors $\bv$.  Since the filtered velocity $\bu=A\bv$, this is equivalent to $\bv\cdot\bu>0$, which means that the global energy (\ref{energy}) is nonnegative and therefore well defined.  A matrix is positive definite if all of its eigenvalues are positive.  Thus a criterion for our filter is that its minimum eigenvalue is positive.  This leads to a process for choosing each filter weight: plot the minimum eigenvalue as a function of the filter weight, and choose the largest weight that guarantees stability.  There is no reason to choose a smaller value, because then the filter smooths less.  Figure \ref{f_pos_evals} shows that this process produces the same restrictions on filter weights as the pressure-velocity instability in the previous section.  In fact, the two methods of analysis agree so well that the pressure-velocity instability appears to be a physical manifestation of the poorly defined energy.  Numerical experiments where the filter weights were varied confirm that the simulations are unstable when the energy is poorly defined.

\section{Results \label{s_results}}

\begin{figure}[tbh]
\center  
\begin{tabular}[c]{cc}
(a)\scalebox{.8}{\includegraphics{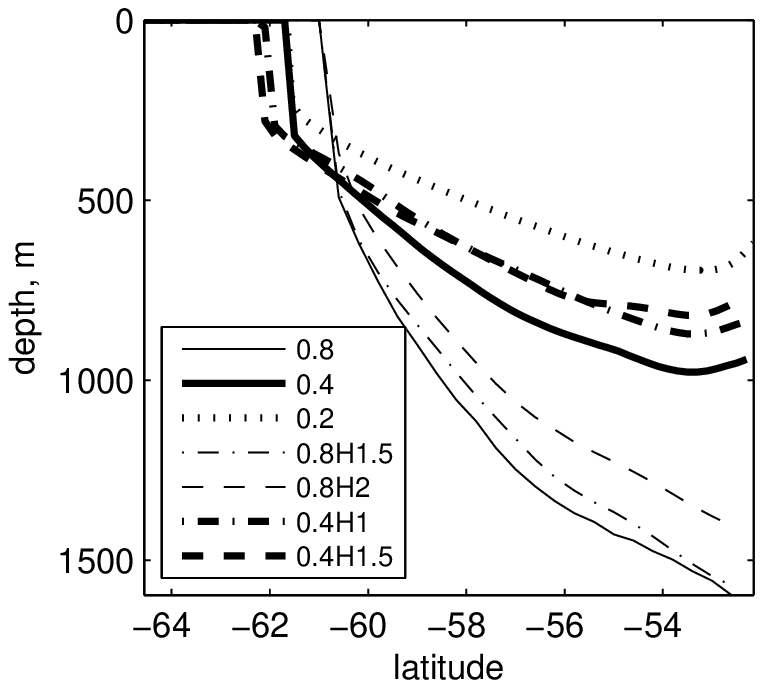}} &
(b)\scalebox{.8}{\includegraphics{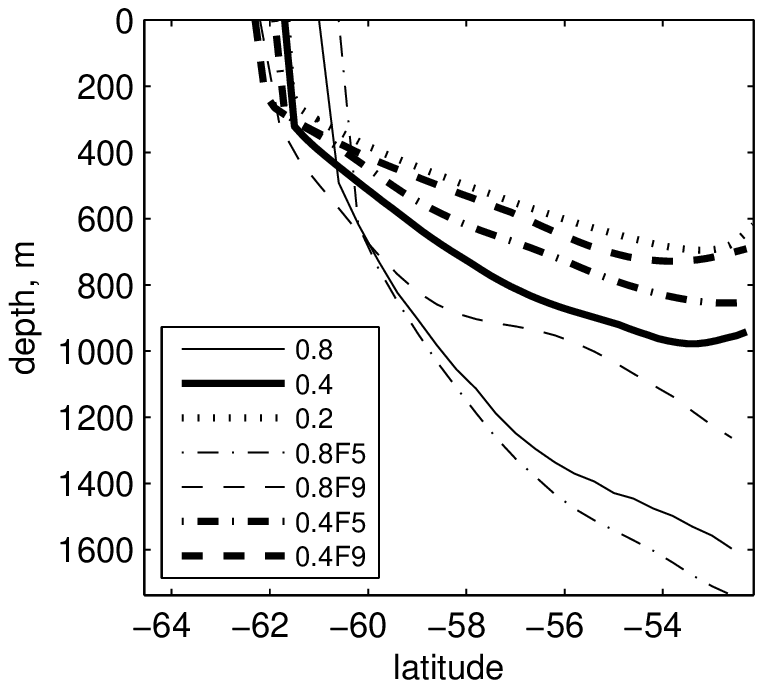}} \\
\end{tabular}
\caption{\label{f_Tdepth6E}
Depth of 6$^o$C isotherm of potential temperature using the Helmholtz inversion (a) and the filter (b) to smooth the velocity.  Higher resolution simulations of standard POP (dotted line) have isotherms which are less tilted than lower resolution simulations (solid lines).  As $\alpha$ or filter size increases, the POP-$\alpha$ simulations show flatter isotherms, and approach a doubling of resolution using standard POP.
}\end{figure}

The model problem is a zonally periodic channel with a deep-sea ridge,
eastward wind forcing, and surface thermal forcing that restores SST
to a smooth profile from 12$^o$C in the north to 2$^o$C in the south,
as described in \cite{Hecht_ea07jcp}.  This configuration is an
idealization of the Antarctic Circulpolar Current, and was designed to
induce baroclinic instability, where isopycnals are tilted from the
horizontal meridionally by the surface thermal and wind forcing.  If
present, eddies in the flow advect heat meridionally, with the net
effect that the isopycnals are less tilted; in this way, the eddies
convert the potential energy of the tilted isopycnals to the kinetic
energy of the eddies themselves.  Simulations of standard POP at three
resolutions show this effect.  In the lowest resolution simulation
(0.8, see Table \ref{t_parameters}) the Rossby Radius of deformation
is not resolved, and no eddies are present.  As the resolution doubles
(0.4) and doubles again (0.2) more eddies exist, and the isopycnals
are less tilted (Fig. \ref{f_Tdepth6E}).

\begin{figure}[tbh]
\center  
\begin{tabular}[c]{ll}
\scalebox{.8}{\includegraphics{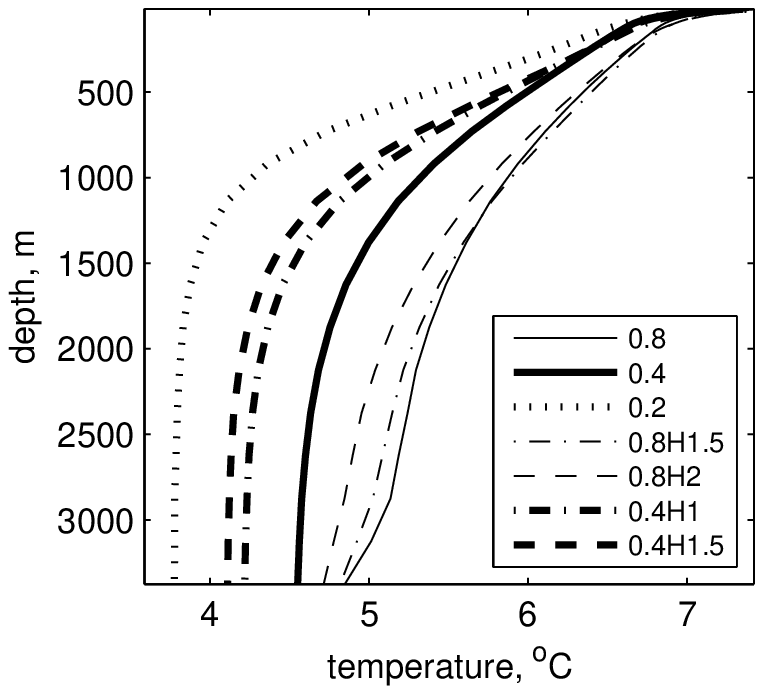}} &
\scalebox{.8}{\includegraphics{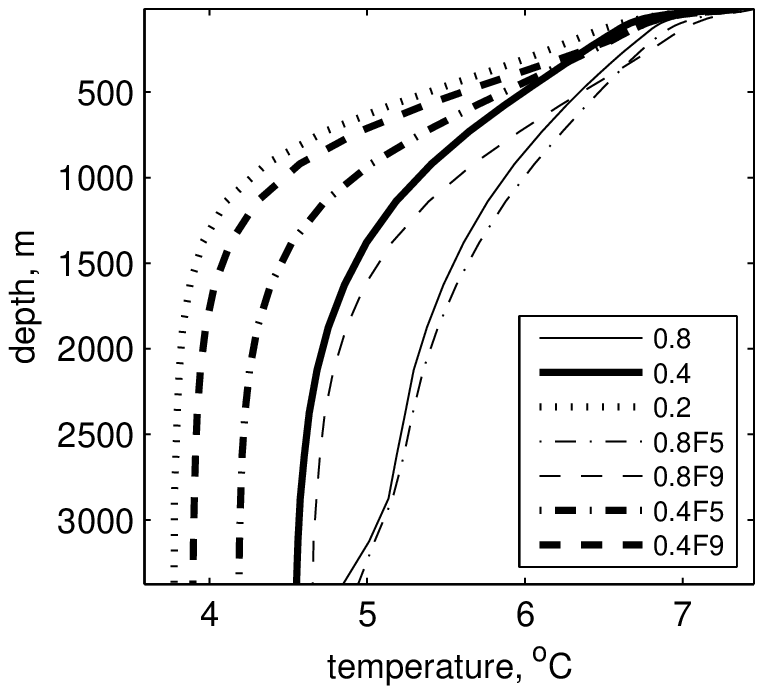}} \\
(a) Helmholtz inversion &
(b) filters
\end{tabular}
\caption{\label{f_temp}
Horizontally averaged potential temperature versus depth using the Helmholtz inversion (a) and the filter (b) to smooth the velocity.  Higher resolution simulations of standard POP (dotted line) are cooler and have a sharper thermocline than lower resolution simulations (solid lines).  Again, POP-$\alpha$ simulations approach a doubling of resolution using standard POP.
}\end{figure}

\begin{figure}[tbh]
\center  
\begin{tabular}[c]{ll}
\scalebox{.7}{\includegraphics{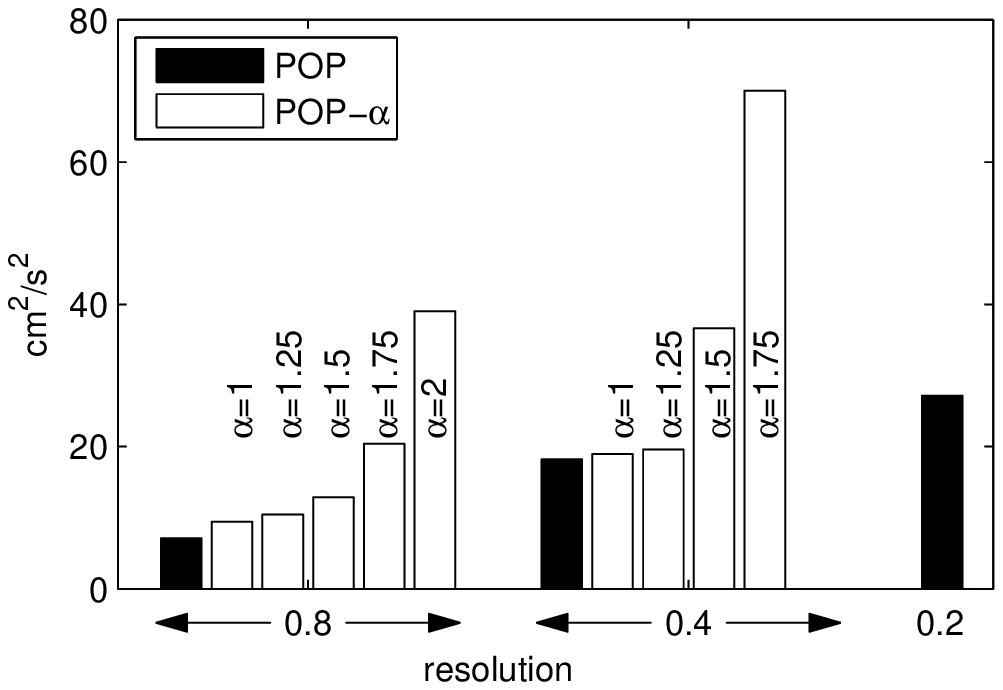}} &
\scalebox{.7}{\includegraphics{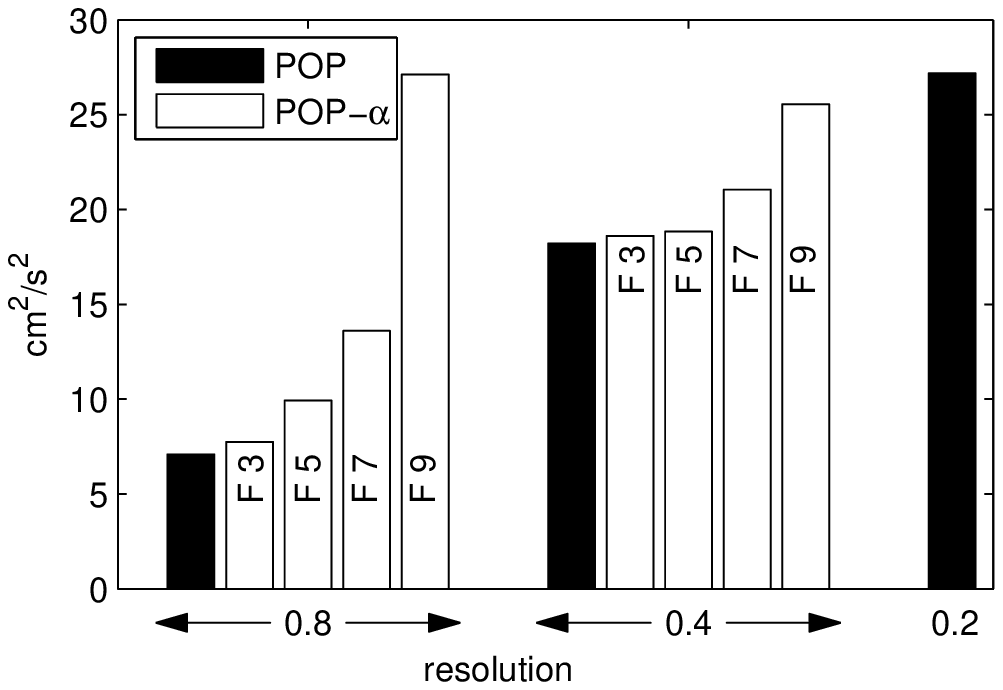}} \\
(a) Helmholtz inversion &
(b) filters
\end{tabular}
\caption{\label{f_KE}
Kinetic energy for all simulations.  As the resolution increases with standard POP, the kinetic energy increases.  Kinetic energy also increases using POP-$\alpha$ at fixed resolution, when $\alpha$ is increased using the Helmholtz inversion (a) or the stencil width of the filter is increased (b).  Each value was calculated by averaging over the entire domain, and in time after 500 years.
}\end{figure}

\begin{figure}[tbh]
\center  
\begin{tabular}[c]{ll}
\scalebox{.7}{\includegraphics{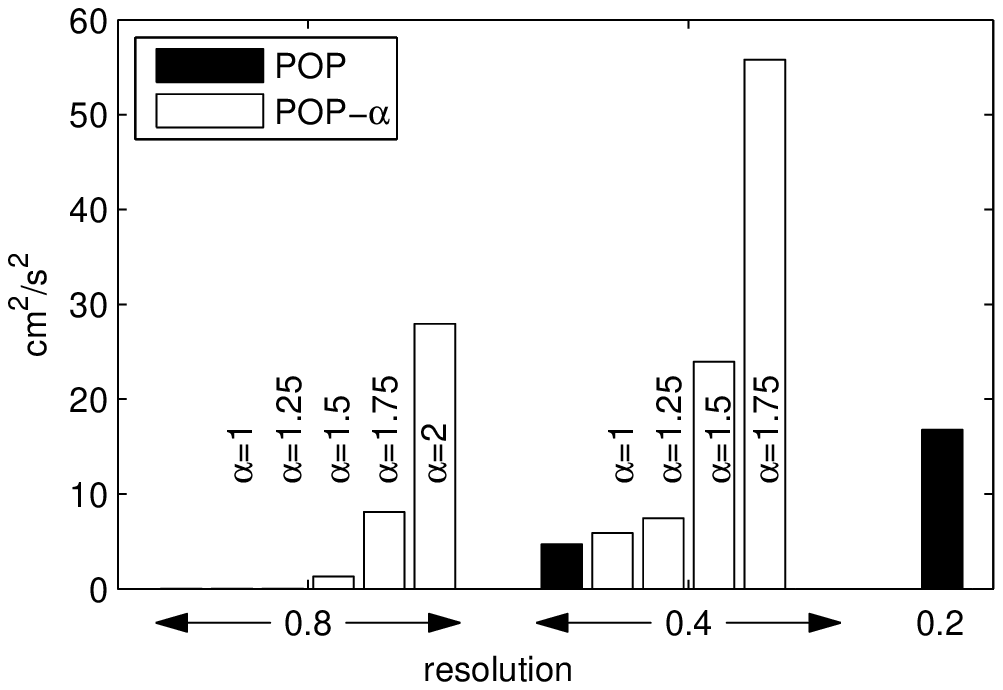}} &
\scalebox{.7}{\includegraphics{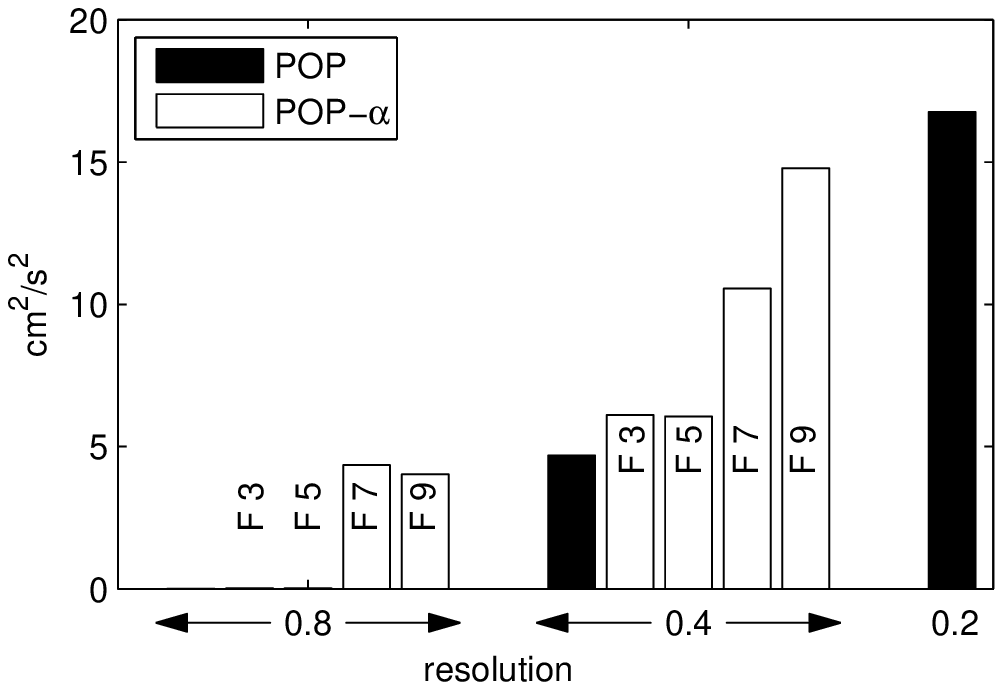}} \\
(a) Helmholtz inversion &
(b) filters
\end{tabular}
\caption{\label{f_EKE}
Same as Fig. \ref{f_KE} but for eddy kinetic energy.  Note that at the lowest resolution (0.8) the eddy kinetic energy of standard POP is indistinguishable from zero on this scale.
}\end{figure}

The flatter isotherms at higher resolution create a sharper thermocline and colder temperatures below the thermocline (Fig. \ref{f_temp}).  (In this model configuration, salinity is constant, so the isotherms shown in the figures coincide with isopycnals.)  Higher resolution simulations of standard POP also show progressively higher kinetic energy (Fig. \ref{f_KE}) and higher eddy kinetic energy (Fig. \ref{f_EKE}) 
as more eddies are resolved.

POP-$\alpha$ simulations resemble higher resolution simulations of
standard POP in all of these measures.  The strength of the eddy
activity is controlled by how much smoothing is done on the rough
velocity $\bv$: when a Helmholtz inversion is used, a larger $\alpha$
smooths more; when a filter is used, wider filter stencils smooth
more.  In either case, Figs. \ref{f_Tdepth6E}--\ref{f_EKE} show that
POP-$\alpha$ can produce results that look progressively like a
doubling of resolution of standard POP as the smoothing is increased.

Because POP-$\alpha$ has two velocities, some details are necessary to explain how these diagnostic quantities were computed.  The kinetic energy is
\bea
KE = \left(u_1^2+u_2^2\right)/2,\;\;\;
KE_\alpha = \left(u_1v_1+u_2v_2\right)/2
\eea
for POP and POP-$\alpha$, where the subscripts indicate horizontal components.
The eddy kinetic energy is
\bea
EKE = \left(\overline{(u_1')^2}+\overline{(u_2')^2}\right)/2,\;\;\;
EKE_\alpha = \left(\overline{u_1'v_1'}+\overline{u_2'v_2'}\right)/2,
\eea
where each variable is the sum of a mean and perturbation, e.g. $u_1=\overline{u_1}+u_1'$, and the overbar indicates a five-year time-averaged mean.  

There would seem to be three possible ways one could compute the kinetic energy for POP-$\alpha$: using the products $u^2$, $uv$, or $v^2$.  The product of the smooth and rough velocities, $uv$, is used because this the conserved kinetic energy that emerges from the derivation of the LANS-$\alpha$ equation \cite{Holm_ea98inbk,Holm99pd}.  For comparison, kinetic and eddy kinetic energies were also computed using $u^2$ and $v^2$.  The product of the smooth velocities is smaller than $uv$, and the product of the rough velocities is larger (typically by 10 to 20\%).  However, the general trends shown in Figs. \ref{f_KE} and \ref{f_EKE} still hold if $u^2$ or $v^2$ are used instead of $uv$.  

An interesting difference between the filter and Helmholtz inversion in POP-$\alpha$ is that simulations using filters are better at the temperature statistics, while simulations using the Helmholtz inversion have higher kinetic and eddy kinetic energy.  This can be seen by comparing 0.8H2 with 0.8F9 or 0.4H1.5 with 0.4F9 if figures \ref{f_Tdepth6E}---\ref{f_EKE}.  Even though results differ a bit with each measure, one may loosely assign each filter an effective alpha value.  For example, in most plots the filter F5 corresponds to $\alpha=1.5\Delta x$ and F9 corresponds to $\alpha=2\Delta x$.

\begin{figure}[tbh]
\scalebox{1}{\includegraphics{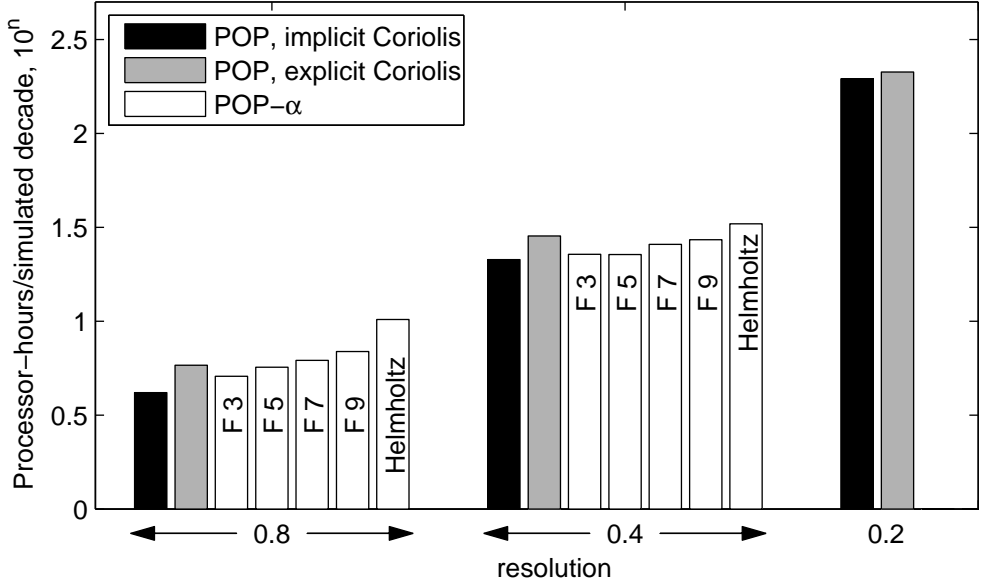}}
\caption{\label{f_timing} 
Timing data from various implementations of POP and POP-$\alpha$ (note log scale).  
The POP-$\alpha$ filter algorithms (F3---F9) are much cheaper than a doubling of resulution using standard POP.  Using a Helmholtz inversion to smooth is more expensive than the filters.
All POP-$\alpha$ algorithms use explicit Coriolis discretization.
}\end{figure}

The computation time required for POP-$\alpha$ is slightly longer than standard POP due to the smoothing operations and additional nonlinear term, but is much less than a doubling of resolution in standard POP (Fig. \ref{f_timing}).  Computation increases as the filter width increases from three to nine; smoothing using the Helmholtz inversion is even slower than the nine-wide filter.  

The standard POP algorithm may use an explicit or implicit discretization for the Coriolis term.  Typically the implicit discretization is chosen because it allows longer timesteps to be taken, resulting in faster simulations.  As described in \cite{Hecht_ea07jcp}, the POP-$\alpha$ algorithm must use the explicit implementation of the Coriolis term.  Fortunately, POP-$\alpha$ can use longer timesteps than the standard POP algorithm with explicit Coriolis \cite{Hecht_ea07jcp,Wingate04mwr}.  This explains why most of the POP-$\alpha$ simulations in Fig. \ref{f_timing} are faster than POP with an explicit Coriolis term, and just a bit slower than POP with an implicit Coriolis term. 

\section{Conclusions \label{s_conclusions}}
This paper presents an assessment of two methods of smoothing the velocity field in the LANS-$\alpha$ turbulence model: the Helmholtz inversion and filters.  The LANS-$\alpha$ model equations specify the Helmholtz inversion, but this method requires an iterative conjugate gradient routine for each smoothing.  The filter, which is simply a weighted average of nearby neighbors, results in simulations that are 20 to 50\% faster than those with the Helmholtz inversion.  This disparity is expected to increase with larger domains and more processors, because more iterations are required for convergence in the conjugate gradient method as the problem-size increases (\cite{Golub_VanLoan96bk}, p. 530).  
The savings in computation time provides a strong motivation to choose a local filter over the global Helmholtz inversion. 

Both the Helmholtz inversion and the filters produce results expected of the LANS-$\alpha$ turbulence model: statistics such as temperature profiles, kinetic and eddy kinetic energy resemble higher-resolution simulations of non-alpha simulations.  The parameter $\alpha$ is a length-scale that controls the amount of smoothing, and thus the strength of the turbulence model.  Analogously, as the filter width varies from three to nine, the smoothing increases and the effects of LANS-$\alpha$ are stronger.  For a particular value of $\alpha$ or filter width, the methods perform a bit differently in different metrics: the filter has better temperature profiles (as judged by higher-resolution simulations), while the Helmholtz inversion has higher kinetic and eddy kinetic energy.

The filter is more robust than the Helmholtz inversion.  When $\alpha$ was too high (i.e the smoothing too strong) in the Helmholtz inversion method, the kinetic energy grew without bound and the simulation was unstable.  This is typical of LANS-$\alpha$ models.  Unstable simulations occurred when $\alpha=2.5\Delta x$ and $2\Delta x$ at resolutions of 0.8 and 0.4, respectively.  For the filter, all stencils tested (up to 9x9) were stable, and stronger sub-grid model effects could be obtained with the filter than with the Helmholtz inversion.

The stencil weights must be chosen carefully, otherwise POP-$\alpha$
will be unstable.  If the outer weights are too large, the smooth
velocity will be of the opposite sign as the rough velocity for the
highest wave-number modes.  When this occurs the pressure gradient,
which is calculated from the smooth velocity, is of the wrong sign.
The physically correct relationship between pressure gradient force
and rough velocity is lost, and the mode grows.  The conditions that
bring about this pressure-velocity instability were identified
exactly, and then verified by numerical simulations.  We found that
the first neighboring weight must be less than one-half the central
weight, and each consecutive neighbor must be less than the one
before.  In 2D, this pattern is squared to make an isotropic, stable
stencil.  The full square was required; leaving off the corners to
make diamond-shaped stencils resulted in unstable simulations.

It is important that the filters chosen are actually smoothing
operators, and that the energy is well defined.  Both of these
properties were verified using an eigenvalue analysis of the filter,
discretized as a matrix operator.  In fact, the criterion on filter
weights that guaranteed well-defined energy turned out to be the same
as the weight limits to prevent the pressure-velocity instability.
Thus it appears that the instability is a physical manifestation of
poorly-defined energy in the equation set.

The goal of any turbulence model is to capture the effects of higher
resolution simulations without paying the computational price of
running at that higher resolution.  We have shown that our
implementation of the LANS-$\alpha$ model in a primitive-equation
ocean-climate model accomplishes this goal.  Specifically, statistics
in POP-$\alpha$ simulations such as kinetic energy, eddy kinetic
energy, and temperature profiles are similar to standard POP
simulations with double the resolution.  But the LANS-$\alpha$
simulations only required 27\% more computing time than standard POP,
as compared to a factor of nine increase to double the resolution of
standard POP (for the 0.4 case).

We are currently performing simulations to compare LANS-$\alpha$ with other turbulence models, including constant coefficient hyperviscosity, Gent-McWilliams isopycnal tracer mixing \cite{Gent_McWilliams90jpo}, the Leray model \cite{Holm99pd}, and the Simplified Bardina model \cite{Cao_ea07unpub}.  These results, to be published in the near future, will consider the merits of LANS-$\alpha$ relative to competing models in the context of primitive-equation ocean models.  We also intend to test the model in an ocean-basin or global simulation.

The comparison of the Helmholtz inversion and the filter in this study has shown that the filter is both cheaper and more robust than the Helmholtz inversion, while producing similar results.  Thus our future work with the POP-$\alpha$ model will use local filters in place of the global Helmholtz for the majority of our simulations. 

\section{Acknowledgements}
I thank M. Hecht and B. Wingate for numerous discussions, ideas, and feedback on the manuscript; E. Titi for valuable insight on the important properties of filters; and B. Geurts for helpful conversations on his experience with filters in LES models.  This work was carried out under the auspices of the National Nuclear Security Administration of the U.S. Department of Energy at Los Alamos National Laboratory under Contract No. DE-AC52-06NA25396.


\setlength{\baselineskip}{14pt}

\bibliographystyle{elsart-num}
\bibliography{alpha_model,ocean_modeling,my_pubs}

\newpage

\clearpage
\newcommand{\Dx}[1]{$\Delta x$}
\begin{table}
\center
  \begin{tabular}[c]{l|cccccccc}
   name & model & smoothing & $\alpha$ & fw & grid & lon & lat\\
   \hline
   \hline
   $0.8$ & POP & - & - & - &  40x40x34  & 0.8 & 0.4 \\
   $0.4$ & POP & - & - & - &  80x80x34  & 0.4 & 0.2 \\
   $0.2$ & POP & - & - & - &  160x160x34  & 0.2 & 0.1 \\
   \hline
   $0.8$H1 & POP-$\alpha$ & Helmholtz & 1 \Dx& - &  40x40x34  & 0.8 & 0.4\\
   $0.8$H1.5 & POP-$\alpha$ & Helmholtz & 1.5 \Dx &  - & 40x40x34  & 0.8 & 0.4\\
   $0.8$H2 & POP-$\alpha$ & Helmholtz & 2 \Dx & - &  40x40x34  & 0.8 & 0.4\\
   \hline
   $0.4$H1 & POP-$\alpha$ & Helmholtz & 1  \Dx& - &  80x80x34  & 0.4 & 0.2\\
   $0.4$H1.5 & POP-$\alpha$ & Helmholtz & 1.5 \Dx &  - & 80x80x34  & 0.4 & 0.2\\
   $0.4$H2 & POP-$\alpha$ & Helmholtz & 2  \Dx& - &  80x80x34  & 0.4 & 0.2\\
   \hline
   $0.8$F3 & POP-$\alpha$ & filter & - & 3 & 40x40x34  &0.8 & 0.4\\
   $0.8$F5 & POP-$\alpha$ & filter & - & 5 & 40x40x34  &0.8 & 0.4\\
   $0.8$F7 & POP-$\alpha$ & filter & - & 7 & 40x40x34  &0.8 & 0.4\\
   $0.8$F9 & POP-$\alpha$ & filter & - & 9 & 40x40x34  &0.8 & 0.4\\
   \hline
   $0.4$F3 & POP-$\alpha$ & filter & - & 3 & 80x80x34  & 0.4 & 0.2\\
   $0.4$F5 & POP-$\alpha$ & filter & - & 5 & 80x80x34  & 0.4 & 0.2\\
   $0.4$F7 & POP-$\alpha$ & filter & - & 7 & 80x80x34  & 0.4 & 0.2\\
   $0.4$F9 & POP-$\alpha$ & filter & - & 9 & 80x80x34  & 0.4 & 0.2\\
 \end{tabular}
\caption{\label{t_parameters}
Model parameters for experiments discussed in this paper, {\it fw} is the filter width; {\it grid} is the number of gridpoints in $(x,y,z)$;  {\it lon} is the longitudinal grid-cell width; and {\it lat} is the latitudinal grid-cell width.  The names are a concatenation of the meridional resolution, the type of smoothing, and the value of $\alpha$ or filter width.
} \end{table}

\begin{table}
\center
  \begin{tabular}[c]{cc|ccc|r@{.}lcc}
{\bf algorithm} & {\bf Cor} & \multicolumn{3}{c}{\bf steps/day} & \multicolumn{4}{c}{\bf clock time} \\
\hline
 & & \multicolumn{3}{c}{resolution} & \multicolumn{4}{c}{resolution} \\
 & &   0.8 & 0.4 & 0.2 &   0&8 & 0.4 & 0.2 \\
   \hline 
POP & imp & 12 & 22 & 40 & 4&16 & 21.3 & 195\\
POP & exp & 20 & 32 & 52 & 5&82 & 28.4 & 212\\
POP-$\alpha$ F3 & exp & 18 & 24 &  & 5&09 & 22.7 & \\
POP-$\alpha$ F5 & exp & 18 & 24 &  & 5&68 & 22.7 & \\
POP-$\alpha$ F7 & exp & 18 & 24 &  & 6&18 & 25.6 & \\
POP-$\alpha$ F9 & exp & 18 & 24 &  & 6&89 & 27.1 & \\
POP-$\alpha$ Helm. & exp & 12 & 18 &  & 10&21 & 33.0 & \\
   \hline
 \end{tabular}
\caption{\label{t_timing}
Minimimum steps/day and the resulting clock time for various algorithms and resolutions.  The second column states whether the barotropic Coriolis term is implicit or exlpicit.  Clock time is in processor-hours per simulated decade.  The fastest POP-$\alpha$ algorithm is the reduced algorithm with a filter.  Even though the POP-$\alpha$ algorithms use explicit Coriolis discretization, the timestep is larger (fewer steps per day) than standard POP with explicit Coriolis.
} \end{table}



\end{document}